\documentclass{article}

\usepackage{PRIMEarxiv}

\usepackage[utf8]{inputenc} % allow utf-8 input
\usepackage[T1]{fontenc}    % use 8-bit T1 fonts
\usepackage{url}            % simple URL typesetting
\usepackage{booktabs}       % professional-quality tables
\usepackage{amsmath}        % math
\usepackage{amsfonts}       % blackboard math symbols
\usepackage{amssymb}        % extra math symbols
\usepackage{nicefrac}       % compact symbols for 1/2, etc.
\usepackage{microtype}      % microtypography
\usepackage{tabularx}
\usepackage{fancyhdr}       % header
\usepackage{graphicx}       % graphics
\usepackage{hyperref}       % hyperlinks (load late)
\hypersetup{colorlinks=true, linkcolor=black, citecolor=black,
            urlcolor=blue, breaklinks=true}
\graphicspath{{./}}         % flat layout: the PNGs sit next to main.tex

\providecommand{\Diamond}{\lozenge}

\title{Characterising Per-Pixel Rendering Difficulty\\
       by Transport Mechanism}

\author{
  Po-Ting Lin \\
  Independent Researcher \\
  Tainan, Taiwan \\
  \texttt{botimlin@gmail.com} \\
}

\begin{document}
\maketitle

% -------------------------------------------------------------------------
\begin{abstract}
Per-pixel rendering difficulty is conventionally characterised by one
noisy scalar: the sample variance of a Monte Carlo estimator. We argue
that it should instead be characterised through \emph{transport
structure} --- a discrete description of how each contribution's energy
reaches the sensor, deterministic under stated renderer conventions ---
with variance treated as a measurement whose reliability that structure
helps predict. We make this concrete with a seven-class
transport-mechanism descriptor assigned per contribution event, evaluate
it on eleven scenes, and measure variance reliability on the seven
first-party ones. The dominant label agrees $87$--$99.6\%$ between 64 and
4096 samples per pixel, where quantile-binned variance agrees as little
as $21\%$; its stability on unseen scenes is predicted from their
pilots. Conditioning a pilot variance on the label improves equal-budget
sample allocation wherever heavy-tailed buckets carry appreciable
population, reduces to the incumbent where they are absent or
negligible, and is neither reproduced by a random partition nor absorbed
by a median-of-means estimator; where the pilot fails for other reasons,
as on the classical ajar-door scene ($6.8$~dB below uniform), the label
says so from the pilot alone.
\end{abstract}

% keywords (replacing the CGF CCS concepts block)

\vspace{0.5em}

% -------------------------------------------------------------------------
\section{Introduction}
\label{sec:intro}

For thirty-five years the operational signal of adaptive Monte Carlo
rendering has been one scalar: the per-pixel sample variance
$\hat\sigma^2(p)$ of an unbiased pixel estimator. Adaptive samplers
refine on it, denoisers steer their filter footprint by it,
efficiency-aware Russian roulette branches off it, and
inverse-rendering pipelines differentiate through it. It is also,
implicitly, the \emph{reference}: the quantity difficulty predictors
are trained against and evaluated on. The assumption underneath is
that a variance estimated from a pilot render is a reliable
measurement of how hard each pixel is.

% -------------------------------------------------------------------------
% Teaser (was \teaser{} in the CGF class; an ordinary figure here).
% Declared after the opening paragraph so it lands at the top of p.2
% rather than above the title.
\begin{figure}[t]
  \centering
  \includegraphics[width=\linewidth]{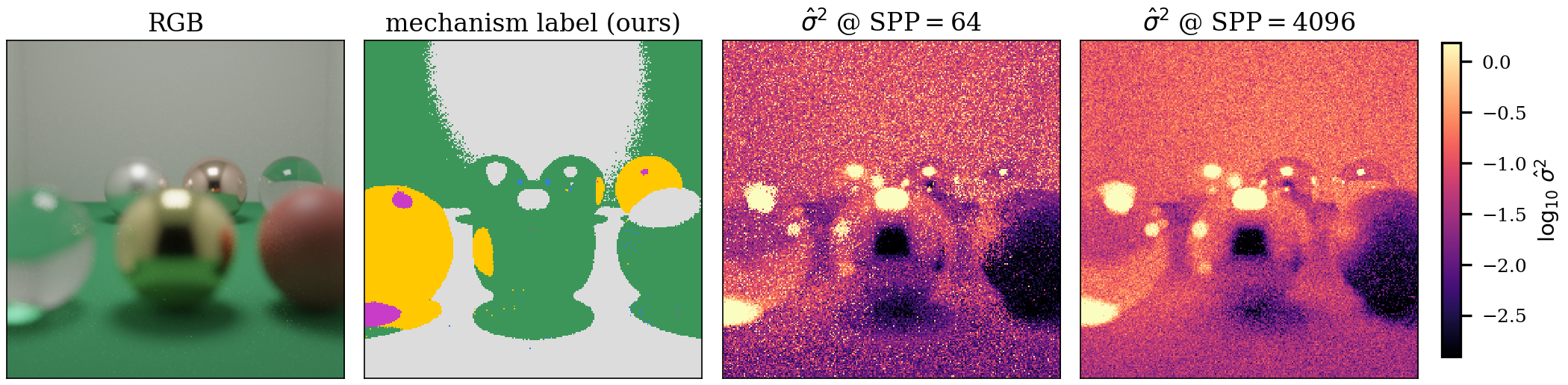}
  \caption{The discrete mechanism label (second panel) is determined
    at SPP$=64$ and barely changes at SPP$=4096$; the residual
    disagreement sits on the boundary between two light-tailed buckets
    (\S\ref{sec:reliable}). The per-pixel sample variance (right two
    panels) drifts substantially across the same $64\times$ budget
    change. \emph{Under heavy-tailed transport, finite-sample variance
    is an unstable measurement; the transport mechanism of each
    contribution is a stable structure --- deterministic under stated
    renderer conventions --- to characterise difficulty by, and to
    condition on.}}
  \label{fig:teaser}
\end{figure}
% -------------------------------------------------------------------------

This paper makes a methodological proposal and supports it with
measurements. The proposal is that per-pixel difficulty should be
characterised through \emph{transport structure} --- the discrete
mechanism by which each contribution's energy reaches the sensor,
deterministic under stated renderer conventions --- and not only
through a noisy continuous statistic. Variance does not disappear
from this picture; it becomes a measurement whose reliability is
itself to be measured, and which the structural description helps
predict. The measurements show three things. First, the reliability
of variance-derived difficulty references is low exactly where
difficulty concentrates, and the failure is associated with specific
transport mechanisms. Second, a mechanism descriptor cheap enough to
read from any path tracer is stable where variance is not
(Fig.~\ref{fig:teaser}), and its
own reliability on three third-party scenes was predicted from their
pilots. Third, conditioning the variance estimate on the descriptor
repairs it where it is least reliable, with an effect a random
partition does not reproduce and a median-of-means estimator does not
absorb.

\paragraph{Why variance is an unreliable reference under heavy tails.}
Three distinct quantities are routinely conflated: the variance
$\sigma^2$ of individual per-sample contributions, the variance
$\sigma^2/N$ of the $N$-sample pixel mean, and the sampling
uncertainty of the \emph{estimator} $\hat\sigma^2_N$. It is the third
that matters for a reference signal. Under finite-fourth-moment
assumptions, the relative standard deviation of the sample-variance
estimator satisfies
$\sqrt{\mathrm{Var}(\hat\sigma^2_N)}/\sigma^2 \approx
\sqrt{(\kappa-1)/N}$, where $\kappa$ is the integrand's fourth
standardised moment \cite{casella2002,vandervaart1998}. Three regimes
follow. When $\kappa$ is finite but large --- caustic and
near-specular integrands --- pilot-budget estimates are useless in
practice. When $\kappa=\infty$ with $\sigma^2$ finite,
$\hat\sigma^2_N$ remains consistent but admits no finite-rate
Gaussian limit \cite{embrechts1997}, so no error bar of the usual
form exists. When $\sigma^2$ itself is infinite --- unbounded
contributions near delta paths \cite{veach1997} --- the target of the
estimate does not exist. Estimating $\kappa$ from a pilot to budget
accordingly fails by regress: it needs the eighth moment. The
rendering literature knows this pathology under other names (weak
singularities \cite{kollig2006}, infinite Hardy--Krause variation
\cite{owen2006}) and suppresses it at the \emph{mean} estimator
\cite{decoro2010,zirr2018}; its implication for $\hat\sigma^2$ as a
reference has not been examined.

The implication is measurable. Rank-correlating per-pixel
$\hat\sigma^2$ between a 64-SPP pilot and a 4096-SPP render gives
Spearman $\rho = 0.26$ on a caustic-dominated scene
(\S\ref{sec:reveal-reliability}). Splitting ten independent seeds
into two halves and fitting a per-pixel convergence exponent on each
--- $\sim$40,000 samples per pixel in total --- gives a split-half
correlation of only $0.23$--$0.29$ on the two geometrically simplest
scenes of our test set. In the language of classical measurement
theory \cite{spearman1904proof,spearman1910correlation} this is a
\emph{reliability} measurement: two independent measurements of the
same target barely agree, and any correlation between that target and
a predictor is attenuated accordingly. We report it as an empirical
reliability figure; turning it into a formal ceiling on predictor
performance would require measurement-error assumptions we do not
need, since low reliability of the evaluation target is damaging
under any of them. Robust estimators of mean and variance under heavy
tails exist \cite{catoni2012challenging,devroye2016subgaussian,lugosi2019mean}
and qualify this critique; \S\ref{sec:related} discusses them and
\S\ref{sec:buys} tests one.

One objection deserves a direct answer. Where $\sigma^2_\infty$
exists, the $\sigma^2$-field is a deterministic functional of the
integrand; the stored buffer $\hat\sigma^2_N/N$ is unbiased for
$\sigma^2_\infty/N$, rank correlations are invariant to the global
$1/N$ factor, and cross-SPP experiments reporting near-unity
\emph{shape} stability in this regime are not in error
\cite{yan2024differentiating,firmino2023denoising}. What is not stable
is the per-pixel \emph{rank} on a scene that contains heavy-tailed
pixels, because those pixels do not contract uniformly with $N$. A
difficulty reference must rank pixels consistently across the entire
scene, including exactly the pixels where $\sigma^2_\infty$ fails to
exist or is recoverable only through fourth-moment-limited estimators.

\paragraph{What we propose instead --- and what we do not claim.}
Three notions must be kept apart. \emph{Rendering difficulty} is
operational and admits several non-equivalent definitions --- variance
under a fixed estimator, samples to an error threshold, time to
quality; we measure the first and third and claim no single scalar
ordering. \emph{Transport mechanism} is structural: how a
contribution's energy reaches the sensor. A mechanism class does not
by itself order pixels by cost --- our own results show large cost
heterogeneity within classes (\S\ref{sec:reveal-cost}) --- so it is
not a difficulty measure; it is an explanatory and \emph{conditioning}
variable for one. \emph{Stability} is what makes a signal usable as a
reference, and proves neither of the other two. Our claim, in this
vocabulary, is that the community conditions on an unstable
measurement when a stable structural descriptor of the same pixels is
available for free. The descriptor we use is the transport mechanism
of each contribution event --- the BSDF lobe at its end-vertex,
whether the path to that vertex passes through a delta interface, and
a single- versus multi-bounce distinction --- which yields seven
mutually exclusive labels assigned from quantities the path tracer
already maintains (\S\ref{sec:descriptor}). The per-event label is
deterministic; the per-pixel label is an estimator of a well-defined
population quantity, and we measure how well it recovers it.

\paragraph{Contributions.}
The paper offers three things that travel beyond its test scenes, and
the findings they produce here.
\begin{enumerate}
\item \textbf{A reliability test for difficulty references.}
      Split-half reliability and cross-budget rank agreement, applied
      per mechanism bucket, audit any per-pixel difficulty signal.
      Applied to variance, they show reliability that is
      tail-dependent and localises to specific mechanisms
      (\S\ref{sec:reveal-reliability}).
\item \textbf{A stable coordinate system.} The seven-class descriptor
      with continuous sidefields, formally separated into per-event
      label, population per-pixel label, and finite-sample estimator.
      On eleven scenes the dominant label agrees at $87$--$99.6\%$
      across a $64\times$ budget change against $21$--$34\%$ for
      7-bin quantile-discretised variance; it survives a controlled
      estimator perturbation; and on three third-party scenes its
      agreement was predicted from each scene's pilot to within
      $0.018$ (\S\ref{sec:reliable}).
\item \textbf{A recommendation with evidence.} Conditioning a pilot
      variance on the label improves equal-budget sample allocation
      wherever heavy-tailed buckets carry appreciable population,
      reduces to the incumbent where they are absent (exactly) or
      negligible (to within $0.004$~dB), is not reproduced by a
      random partition, and is
      not absorbed by a median-of-means variance estimator
      (\S\ref{sec:buys}).
\end{enumerate}
Read per mechanism, difficulty shows structure a scalar hides: in our
scenes caustic transport is not the most expensive --- narrow-lobe
glossy and deep specular chains are, and time-to-quality is
right-censored there (\S\ref{sec:reveal-cost}); lobe width
monotonically governs the glossy firefly tail
(\S\ref{sec:reveal-lobe}); the coupling between delta-mediated and
glossy energy follows mutual visibility between object families, and
held as a pre-registered prediction on one third-party scene
(\S\ref{sec:structure-visibility}); and a scene can be heavy-tailed
in variance yet light-tailed in mechanism, which the label indicates
from the pilot alone (\S\ref{sec:buys-ajar}).

% -------------------------------------------------------------------------
\section{Related Work}
\label{sec:related}

Three bodies of work produce or rely on a per-pixel quantity treated
as a difficulty signal: sample-statistic estimates of variance,
analytically traced continuous descriptors, and path mechanism used
inside the estimator or for compositing.

\subsection{Variance as the de facto difficulty signal}
\label{sec:related-variance}

Image-space adaptive sampling refines where a perceptually motivated
contrast test fires \cite{mitchell1987} or greedily minimises
per-pixel relative MSE \cite{rousselle2011}; variance-guided denoising
builds its filter footprint from a variance buffer \cite{schied2017};
efficiency-aware path tracing iterates over second-moment estimates
\cite{rath2022}; Bayesian direct illumination regresses over per-light
statistics \cite{vevoda2018}; the Eurographics STAR \cite{zwicker2015}
surveys the family as the dominant paradigm. Inverse rendering derives
unbiased estimators for the derivatives of variance
\cite{yan2024differentiating}, and denoising-aware adaptive sampling
drives its sampling map from a variance estimate of the denoiser's
output \cite{firmino2023denoising}. The community has noticed that
$\hat\sigma^2$ behaves badly under firefly-prone transport ---
density-based outlier rejection \cite{decoro2010}, firefly
re-weighting \cite{zirr2018}, and the observation that variance and
contrast cannot be reliably computed at low sample counts
\cite{kuznetsov2018} --- but in each case the instability is
suppressed at the mean estimator rather than examined as a property
of the reference.

\paragraph{Robust estimation qualifies the critique.}
Estimators of mean and variance with sub-Gaussian deviation
guarantees under heavy tails --- Catoni's deviation study
\cite{catoni2012challenging}, sub-Gaussian mean estimators
\cite{devroye2016subgaussian}, the median-of-means family
\cite{lugosi2019mean} --- need only the moments their bounds assume,
not the finite fourth moment that error bars on $\hat\sigma^2$
require; within graphics, recent work corrects estimated radiance and
variance from statistical models of the sample population
\cite{sakai2025stater}. These methods weaken the practical force of
the instability wherever their moment assumptions hold. They do not
remove the regime structure --- when the fourth moment is unbounded,
variance estimation degrades regardless of estimator --- nor supply
mechanism-level structure. We treat robust estimation and mechanism
conditioning as complementary and test the relationship for one
such estimator, median-of-means (\S\ref{sec:buys-robust}). Per-pixel guidance maps produced by
denoising pipelines \cite{bako2017kpcn,vogels2018denoising} inherit
the reliability of their variance inputs at low sample counts; the
split-half methodology of \S\ref{sec:reveal-reliability} applies to
them directly, though quantifying it is beyond our scope.

\subsection{A-priori prediction via traced continuous descriptors}
\label{sec:related-covariance}

A second line forgoes sample variance and predicts difficulty
analytically by tracing a continuous descriptor of the radiance
signal through the transport operators: frequency analysis
\cite{durand2005}, its specialisation to depth of field
\cite{soler2009}, the $5{\times}5$ covariance matrix propagated
through transport, BRDF, occlusion, lens and motion
\cite{belcour2013covariance}, and first-order sensitivities of
lighting, shading and shadow \cite{ramamoorthi2007}. These are the
closest prior work by goal. Their object --- a covariance, spectrum or
gradient field --- is itself continuous and must be traced and
approximated under Gaussian-family assumptions; our label is a
deterministic function of path structure read from a single
contribution event.

\subsection{Path mechanism inside the estimator}
\label{sec:related-mechanism}

Path mechanism has been load-bearing for decades, but inside the
estimator or as a compositing convenience. Heckbert's
\texttt{L(D|S)*E} notation \cite{heckbert1990} described paths and
motivated hybrid algorithms; Veach \cite{veach1997} formalised it as
a classifier of which algorithms can sample which configurations.
Photon mapping separates a caustic from a global map \cite{jensen1996};
path guiding learns incident-radiance proposals as parametric
mixtures \cite{vorba2014}, SD-trees \cite{muller2017practical} or
normalising flows \cite{muller2019neural}; manifold next-event
estimation \cite{hanika2015} and specular manifold sampling
\cite{zeltner2020} target specific mechanisms. Path-space
regularization makes the mechanism--difficulty connection explicit
from the estimator side, mollifying unsampleable or high-variance
path classes by path-space criteria
\cite{kaplanyan2013regularization} or by widening microfacet lobes
\cite{jendersie2019microfacet} --- direct evidence that particular
mechanisms are trouble spots for particular estimators, though
neither produces a per-pixel label, and estimator-specific trouble
does not by itself imply an estimator-independent ordering of
difficulty. Variance-aware MIS \cite{grittmann2019} combines
mechanism and variance at sampler-family granularity. Production
light-path expressions \cite{rendermanLPE} are the closest analogue
to our taxonomy --- a regex over \texttt{L/D/G/S/E} selecting events
into AOVs --- but are user-defined, overlapping, and framed as
compositing conveniences. The mechanism abstraction has appeared in
every role except as a per-pixel descriptor of, and conditioning
signal for, rendering difficulty. That is the role proposed here.

% -------------------------------------------------------------------------
\section{The Transport-Mechanism Descriptor}
\label{sec:descriptor}

We describe each pixel by a discrete label assigned per contribution
event. The \emph{per-event} label is a deterministic function of the
lobe sampled at the event's end-vertex and the lobe sequence leading
to it, under the estimator conventions of \S\ref{sec:descriptor-scope};
nothing at the event level is estimated or thresholded. The
\emph{per-pixel} label aggregates events by an energy-weighted argmax
and is therefore a Monte Carlo estimate at finite sample counts.

\subsection{The contribution event}
\label{sec:descriptor-event}

Each contribution to a pixel value reaches the sensor through a
specific sequence of vertices and BSDF events; we call each
accumulation a \emph{contribution event}. In a path tracer with
next-event estimation they arise in three forms: a successful NEE
shadow ray, a BSDF-sampled bounce that lands on an emitter, and the
trivial primary-ray-into-emitter case. For each event we record
(a)~the BSDF lobe at the \emph{end-vertex} $V_n$, the last surface
vertex before the emitter; (b)~whether the light-to-end-vertex
sequence contains any delta-specular interaction; and (c)~the event's
surface-bounce count. All three are read from lobe types and counters
the path tracer already maintains; no additional rays are cast.

\paragraph{End-vertex lobe under NEE.}
For a BSDF-sampled bounce the lobe at $V_n$ is the BSDF's reported
sampled lobe. For an NEE connection the BSDF is only evaluated, so
$V_n$ takes a lobe from the BSDF's flags: \textbf{D} if it advertises
a diffuse component, \textbf{G} otherwise; multi-lobe BSDFs
advertising both (e.g.\ \texttt{roughplastic}) are routed to
\textbf{D}, the light-tailed side. Delta surfaces have zero NEE
connection probability, so an NEE end-vertex is never \textbf{S} and
\texttt{specular-direct} receives only BSDF-sampled contributions.

\paragraph{Population label versus finite-sample label.}
Let $L_b(p)$ be the population path-space contribution to pixel $p$
restricted to bucket $b$ --- the integral of the measurement
contribution function over paths whose events classify to $b$. The
\emph{ideal dominant label} $b^*(p) = \arg\max_b L_b(p)$ is a
deterministic property of the integrand and the stated conventions. A
renderer at budget $N$ produces estimates $\hat L_b^{(N)}(p)$, and the
reported label is $\hat b_N(p) = \arg\max_b \hat L_b^{(N)}(p)$: an
estimator of $b^*(p)$, and at finite SPP sample-, estimator- and
implementation-dependent. The stability results of
\S\ref{sec:reliable} are evidence that this estimator recovers
$b^*(p)$ reliably at small budgets in the tested settings; the
practical cost of estimating it from a 4-SPP pilot is measured in
\S\ref{sec:buys} and found to be negligible.

\subsection{The partition}
\label{sec:descriptor-partition}

Each event maps to one of seven buckets (Table~\ref{tab:partition})
along two axes --- end-vertex lobe $\in\{\mathrm{D},\mathrm{G},\mathrm{S}\}$
and presence of a $\delta$-S event in the sequence --- refined within
the no-$\delta$ branch by a single- versus multi-bounce predicate,
with a seventh row for the trivial primary-ray case. Operationally it
is a first-match cascade, with end-vertex lobe $\ell$, delta flag
$\delta$ and bounce count $n$:
(1)~$n=0$: \texttt{emitter-direct};
(2)~$\ell=\mathrm{S}$: \texttt{specular-direct};
(3)~$\ell\notin\{\mathrm{D},\mathrm{G}\}$: \texttt{other};
(4)~$\delta$: \texttt{delta-mediated};
(5)~$n=1$: \texttt{direct};
(6)~$\ell=\mathrm{D}$: \texttt{diffuse-indirect};
(7)~$\ell=\mathrm{G}$: \texttt{glossy}.
The cascade matches the integrator implementation line for line.

\begin{table}[!htbp]
\centering
\small
\caption{The seven contribution-event buckets. \texttt{delta-mediated}
marks a \emph{transport class} --- energy that passed through a delta
interface --- of which the bright image-space caustic footprint is a
subset, resolved by the continuous $\delta$-share sidefield
(\S\ref{sec:descriptor-sidefields}).}
\label{tab:partition}
\begin{tabular}{@{}llll@{}}
\hline
end-vertex lobe & $\delta$-S in seq. & bounces & bucket \\
\hline
(no surface bounce) & --- & $=0$ & \texttt{emitter-direct} \\
D or G & no & $=1$ & \texttt{direct} \\
D & no & $>1$ & \texttt{diffuse-indirect} \\
G & no & $>1$ & \texttt{glossy} \\
D or G & yes & any & \texttt{delta-mediated} \\
S & --- & --- & \texttt{specular-direct} \\
other & --- & --- & \texttt{other} \\
\hline
\end{tabular}
\end{table}

The partition is mutually exclusive and formally complete by
construction --- the \texttt{other} row catches anything outside the
named cells. What construction does not guarantee is that the six
\emph{named} mechanisms cover the transport that actually occurs;
\S\ref{sec:reliable-coverage} measures this. Every cell boundary is a
discrete predicate on a lobe type, a sequence property or the bounce
count; no continuous threshold enters the main axis.

\subsection{Continuous sidefields}
\label{sec:descriptor-sidefields}

The dominant label discards the per-pixel mixture. Four continuous
sidefields retain it: \emph{purity} $\pi(p)\in[0,1]$, the energy
fraction of the dominant bucket; the \emph{glossy energy fraction}
$\gamma(p)$, the share of events whose sequence contains any glossy
bounce; the \emph{energy-weighted bounce depth} $d(p)$; and the
\emph{per-bucket energy fractions} $\hat L_b/\sum_{b'}\hat L_{b'}$, of
which the \emph{$\delta$-share} (the \texttt{delta-mediated}
fraction) resolves the bright visual caustic footprint from pixels
that merely receive some delta-mediated energy. Any threshold in the
framework is confined to sidefields, never to the main partition.

\subsection{Design choices}
\label{sec:descriptor-design}

\paragraph{Why discrete.}
A discrete label absorbs small perturbations --- a design property,
not by itself evidence that the label measures difficulty, since any
coarse quantisation is more stable than the continuous signal it
coarsens. Two things elevate the property beyond quantisation: the
fair comparison of \S\ref{sec:reliable-stability}, in which
$\hat\sigma^2$ discretised onto quantile bins of matching cardinality
is still $1.4$--$4.5\times$ less stable; and the placebo test of
\S\ref{sec:buys}, in which a random partition with identical class
sizes does not reproduce the label's operational value. What matters
is \emph{which} pixels share a class, not that classes exist.

\paragraph{Why $\gamma$ is continuous.}
A one-bit ``any glossy interaction'' flag saturates ($0.0$, $0.0$,
$1.0$ on \texttt{glass\_caustic}, \texttt{cubes\_dof},
\texttt{snooker} at SPP$=$4096) and would saturate at $1.0$ on any
glossy-rich scene; the continuous fraction recovers the distinction
between ``occasionally glossy'' and ``glossy-dominated'' at no cost,
since the path tracer accumulates the glossy-touching stream anyway.

\paragraph{Why glossy is its own bucket.}
Folding glossy focusing into \texttt{delta-mediated} would require a
roughness threshold on the main axis; and the empirical relationship
between the two has no fixed sign --- it reverses with scene geometry
(\S\ref{sec:structure-visibility}), so any rule pre-committing glossy
to either neighbour mislabels at least one observed regime.

\subsection{Scope and conventions}
\label{sec:descriptor-scope}

The descriptor is a property of the integrand \emph{together with}
stated conventions, not of the integrand alone. The NEE lobe rule and
the D/G/S trichotomy depend on the renderer's material model and flag
semantics; a different BSDF decomposition, roughness parameterisation
or layered-material factoring may label the same physical scattering
function differently. Appendix~\ref{sec:appendix} gives the complete
mapping for every material used; rather than assert insensitivity, we
perturb the estimator and report which pixels move
(\S\ref{sec:reliable-robust}). All claims concern surface path tracing
on scenes of diffuse, rough- and smooth-conductor, rough- and
smooth-dielectric, and plastic-family materials. We make no claims for
participating media, subsurface transport, hair and foliage
visibility complexity, glinty or measured materials, environment
emitters, or null scattering --- several of which carry substantial
difficulty regimes of their own (\S\ref{sec:scope}).

% -------------------------------------------------------------------------
\section{Coverage, Stability and Robustness of the Descriptor}
\label{sec:reliable}

Before using the descriptor to characterise anything, we establish
its own reliability: the named mechanisms cover the transport that
occurs on the tested scenes; the per-pixel label is stable across an
order-of-magnitude budget change in the regime where variance-derived
references are not; it survives a controlled perturbation of the
estimator; and on three third-party scenes its stability was
predicted from each scene's own pilot.

\subsection{Test matrix}
\label{sec:reliable-matrix}

Seven first-party scenes span the regime of interest:
\texttt{cubes\_dof} (three cubes, thin lens, Lambertian only);
\texttt{glass\_caustic} (Cornell variant, glass sphere, L-S-D
caustic on the floor); \texttt{snooker} (six balls --- matte plastic,
polished gold/silver/copper, glass --- on felt);
\texttt{tabletop\_mixed}/\texttt{tabletop\_separated} (a controlled
pair sharing geometry and materials, the second adding an opaque
partition that blocks delta-S$\leftrightarrow$glossy sight lines);
\texttt{ring\_caustic} (two polished metal rings on a diffuse table
under a small bright light --- a lying ring focusing the classic
cardioid, a standing ring projecting a caustic streak; every caustic
path is L\,S$^+$\,D\,E, giving an unambiguous, visually verifiable
delta-mediated population of $6{,}949$ pixels); and
\texttt{kitchen\_counter} (${\sim}15$ objects spanning diffuse,
rough plastic, brushed and polished conductors at several
roughnesses, smooth glass and a frosted tumbler, two lights; metal
objects before a brushed-steel backsplash give the \texttt{glossy}
bucket $10{,}137$ pixels). Four third-party scenes from Bitterli's
collection \cite{bitterli2016resources} are used in their native
Mitsuba~3 form --- \texttt{veach-ajar}, \texttt{veach-mis},
\texttt{bathroom2}, \texttt{matpreview} --- each chosen to test one
claim on geometry the authors did not design
(\S\ref{sec:reliable-transfer}, \S\ref{sec:buys-ajar},
\S\ref{sec:scope}).

Each first-party scene is rendered at SPP $\in\{64,512,4096\}$; the
SPP$=$4096 render is the reference and the partition is computed at
each SPP independently. The implementation is a Mitsuba~3 custom
\texttt{SamplingIntegrator} recording, per contribution event, the
end-vertex lobe, a delta-in-sequence bit and the bounce count. All
tabulated matrices use a single renderer version, and re-rendering
the original test matrix from scratch reproduced every statistic to
within seed noise; four auxiliary measurements --- the cross-budget
convergence study behind $\rho=0.26$, the $\hat\alpha$ reliability
figure, the \texttt{snooker} glossy sub-population seed statistics,
and the light-count control --- date from the original runs and are
used unchanged (Appendix~\ref{sec:appendix}).

\subsection{Coverage}
\label{sec:reliable-coverage}

\begin{figure*}[!t]
  \centering
  \includegraphics[width=\textwidth]{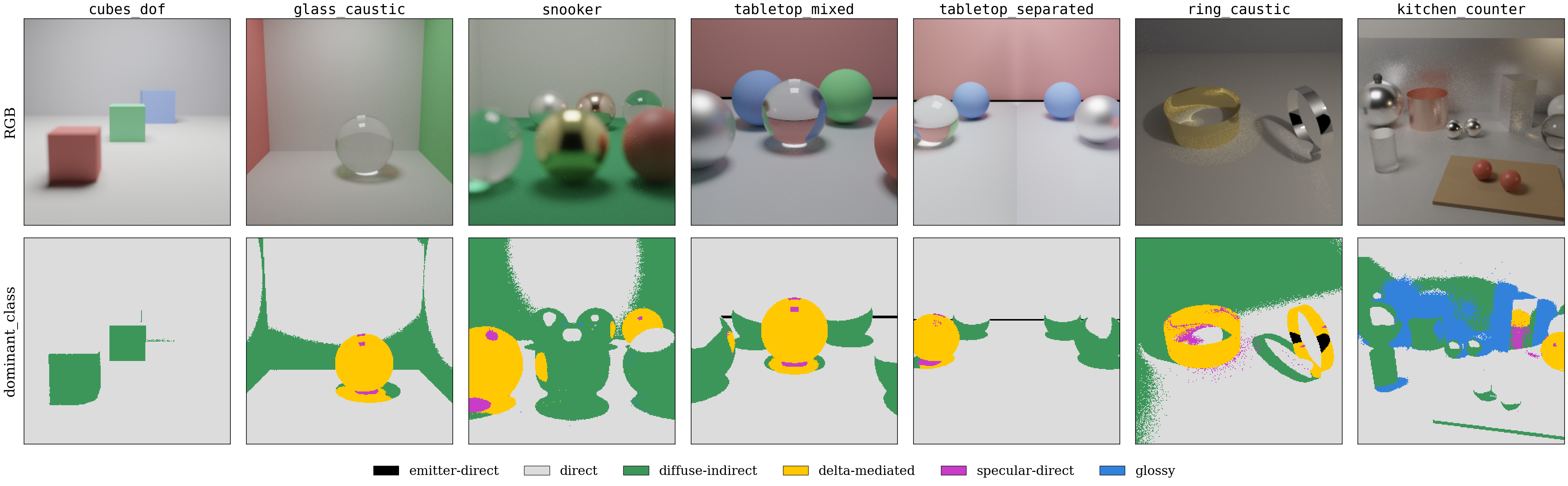}
  \caption{RGB renders (top) and dominant-label maps (bottom) for the
    seven first-party scenes. The tabletop pair differs only by the
    blocking partition; \texttt{ring\_caustic}'s lying-ring cardioid
    is directly visible in the RGB render; \texttt{kitchen\_counter}'s
    metal cluster and brushed backsplash form the $10{,}137$-pixel
    \texttt{glossy} population used for the population-scale analyses
    of \S\ref{sec:reliable-robust} and \S\ref{sec:reveal-cost}.}
  \label{fig:buckets}
\end{figure*}

The \texttt{other} catch-all receives $0\%$ of contribution-event
energy on all 21 first-party runs and on every third-party render
(both budgets, both estimator modes, all material variants): on the
scenes tested, the six named mechanisms account for all observed
energy. This establishes coverage for the tested scene class, not
universal completeness --- materials outside the D/G/S trichotomy
would land in \texttt{other} by design. Per-scene bucket maps agree
with physical expectation (Fig.~\ref{fig:buckets}): \texttt{cubes\_dof}
is $90.9\%$ \texttt{direct}; \texttt{glass\_caustic} places $7.4\%$
of pixels ($10.0\%$ of energy) into \texttt{delta-mediated} on the
floor below the sphere; \texttt{snooker} sorts glass and chrome balls
into \texttt{delta-mediated} and the gold ball into \texttt{glossy};
\texttt{ring\_caustic} concentrates \texttt{delta-mediated} on and
inside the rings; \texttt{kitchen\_counter} assigns its metal cluster
and backsplash reflections to \texttt{glossy}. Such agreement is a
sanity check, not proof of a uniquely correct partition.

\subsection{Cross-budget stability}
\label{sec:reliable-stability}

Table~\ref{tab:cross-spp} reports per-pixel agreement of the dominant
label between budgets and the largest change in any bucket's pixel
share. Raw agreement could be inflated by a majority bucket; Cohen's
$\kappa$ \cite{cohen1960coefficient}, which removes chance agreement,
stays $0.79$--$0.98$ across all seven scenes
(Table~\ref{tab:kappa}), ``substantial'' to ``almost perfect'' on the
Landis--Koch scale \cite{landis1977measurement}, used informally.

\begin{table}[!htbp]
\centering
\small
\caption{Cross-budget stability of the dominant label: per-pixel
agreement between budgets, and the largest change in any bucket's
pixel share between SPP$=64$ and $4096$.}
\label{tab:cross-spp}
\setlength{\tabcolsep}{3.5pt}
\begin{tabular}{@{}lccc@{}}
\hline
scene & 64\,vs\,4096 & 512\,vs\,4096 & max drift \\
\hline
\texttt{cubes\_dof} & 99.58\% & 99.82\% & 0.12\% \\
\texttt{tabletop\_separated} & 99.43\% & 99.77\% & 0.12\% \\
\texttt{tabletop\_mixed} & 98.65\% & 99.45\% & 0.37\% \\
\texttt{glass\_caustic} & 95.64\% & 98.70\% & 1.19\% \\
\texttt{snooker} & 89.55\% & 95.77\% & 0.58\% \\
\texttt{kitchen\_counter} & 88.69\% & 94.12\% & 2.88\% \\
\texttt{ring\_caustic} & 87.07\% & 92.81\% & 1.92\% \\
\hline
\end{tabular}
\end{table}

\begin{table}[!htbp]
\centering
\small
\caption{Same-scale comparison and chance correction (64 vs 4096).
$\hat\sigma^2$ is binned into per-scene quantile bins computed
independently at each SPP --- which removes the global $1/N$
rescaling and scores only the stability of the field's shape, the
reading most favourable to variance --- and scored by the same
agreement metric as the label. Weighted $\kappa$
\cite{cohen1968weighted} uses penalty $0.2$ for swaps within
\{\texttt{emitter-direct}, \texttt{direct}, \texttt{diffuse-indirect}\}
and $1.0$ for any swap touching \{\texttt{delta-mediated},
\texttt{specular-direct}, \texttt{glossy}\}; the classes are not
ordinal, and this weighting is declared as our reading of boundary
flicker versus structural change. ``struct\%'' is the share of
disagreeing pixels involved in a swap touching the second family.}
\label{tab:kappa}
\setlength{\tabcolsep}{2.6pt}
\begin{tabular}{@{}lcccccc@{}}
\hline
scene & $\hat\sigma^2$ 3-bin & 7-bin & label & $\kappa$ & wtd.\ $\kappa$ & struct\% \\
\hline
\texttt{cubes\_dof} & 0.484 & 0.264 & 0.996 & 0.975 & 0.975 & 0.0\% \\
\texttt{tabletop\_sep.} & 0.652 & 0.335 & 0.994 & 0.974 & 0.972 & 47.2\% \\
\texttt{tabletop\_mixed} & 0.596 & 0.286 & 0.987 & 0.964 & 0.953 & 70.3\% \\
\texttt{glass\_caustic} & 0.424 & 0.212 & 0.956 & 0.890 & 0.900 & 32.9\% \\
\texttt{snooker} & 0.560 & 0.307 & 0.896 & 0.828 & 0.885 & 17.0\% \\
\texttt{kitchen\_counter} & 0.611 & 0.344 & 0.887 & 0.789 & 0.718 & 81.3\% \\
\texttt{ring\_caustic} & 0.524 & 0.265 & 0.871 & 0.786 & 0.859 & 14.3\% \\
\hline
\end{tabular}
\end{table}

\paragraph{The fair comparison.}
A coarse categorical variable is naturally more stable than a
continuous one, so the raw juxtaposition ``$95.6\%$ versus
$\rho=0.26$'' (both \texttt{glass\_caustic}) proves little. Discretising $\hat\sigma^2$ onto
quantile bins of matching cardinality and scoring it with the
identical metric (Table~\ref{tab:kappa}, first columns) gives
$0.42$--$0.65$ (3-bin) and $0.21$--$0.34$ (7-bin) against the label's
$0.87$--$0.996$: on every scene the label is $1.4$--$2.3\times$ more
stable than coarse-binned variance and $2.5$--$4.5\times$ more than
fine-binned. Class imbalance is handled by $\kappa$, not raw
agreement.

\begin{figure}[!htbp]
  \centering
  \includegraphics[width=0.9\columnwidth]{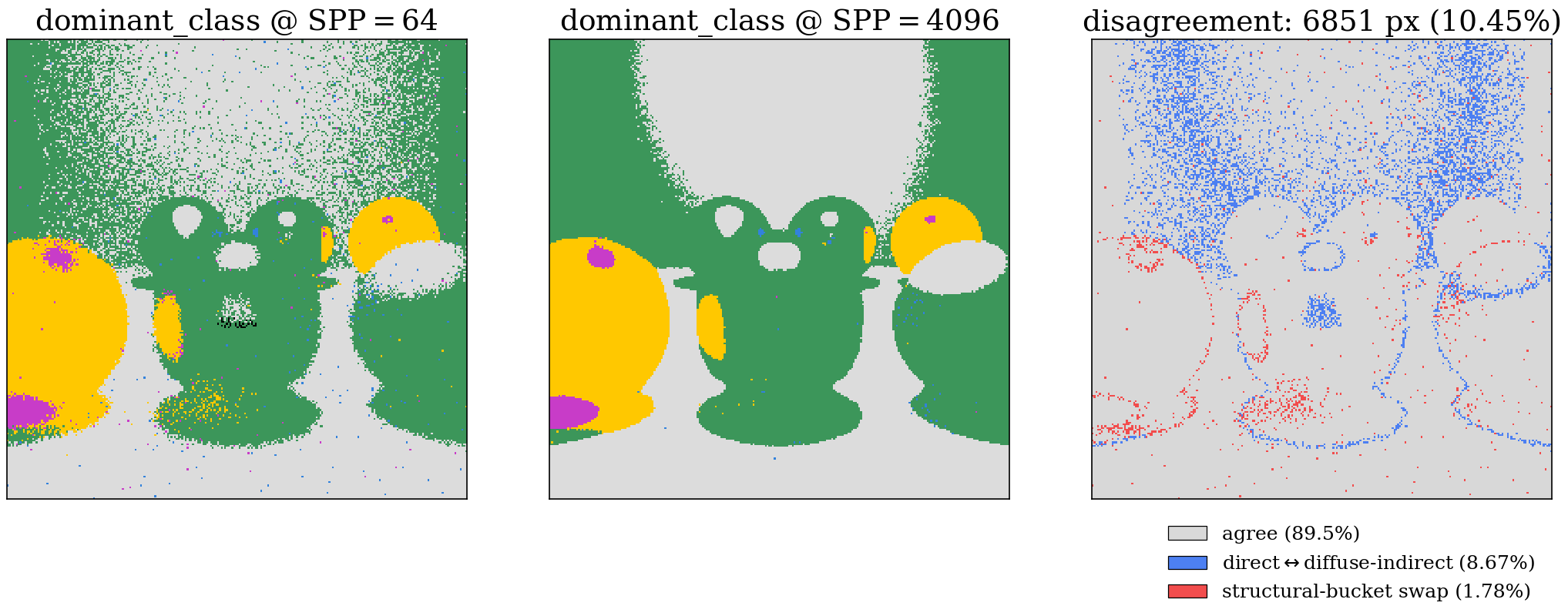}
  \caption{Cross-budget stability on \texttt{snooker}. Left/middle:
    dominant label at SPP$=64$ and $4096$. Right: agreement map (grey
    = agree, blue = \texttt{direct}$\leftrightarrow$\texttt{diffuse-indirect}
    swap, red = swap touching a structural bucket). $89.55\%$ agree;
    the $10.45\%$ disagreement is dominated by the light-tailed
    boundary ($8.67\%$), with $1.78\%$ of pixels in any structural
    swap; no bucket's share moves by more than $0.58\%$.}
  \label{fig:cross-spp}
\end{figure}

\paragraph{Where the residual disagreement lives.}
Two kinds of label boundary must be distinguished. \emph{Predicate
boundaries} are decided by a discrete test --- membership of
\texttt{delta-mediated} and \texttt{specular-direct} turns on a
$\delta$-S event. \emph{Energy-argmax boundaries} are decided by
which bucket carries more energy: \texttt{direct} versus
\texttt{diffuse-indirect} (bounce-count split) and
\texttt{diffuse-indirect} versus \texttt{glossy} (lobe split). Across
all seven scenes disagreement concentrates on the energy-argmax
boundaries and spares the predicate buckets. On the delta-heavy
scenes the dominant flicker is
\texttt{direct}$\leftrightarrow$\texttt{diffuse-indirect}
(Fig.~\ref{fig:cross-spp}), and weighted $\kappa$, which forgives
that swap, rises above Cohen's ($0.828\to0.885$ on \texttt{snooker},
$0.786\to0.859$ on \texttt{ring\_caustic}). On the glossy-rich
\texttt{kitchen\_counter} the dominant flicker is instead
\texttt{diffuse-indirect}$\leftrightarrow$\texttt{glossy} ($2{,}575$
of $7{,}412$ disagreeing pixels at SPP$=$64), which our weighting
scores as structural --- hence the one weighted $\kappa$ that drops
and the $81\%$ struct share. That is the glossy \emph{energy}
boundary, not a predicate failure: \texttt{kitchen\_counter}'s
\texttt{delta-mediated} share drifts by $0.01\%$ across the $64\times$
change while its \texttt{diffuse-indirect}/\texttt{glossy} shares
exchange up to $2.9\%$ of pixels. Structural-bucket shares drift by
at most $2.9\%$ anywhere, and by $\le1.2\%$ outside that scene. The
continuous sidefields drift as expected for continuous quantities:
purity median by $0.94$--$2.4\%$, glossy fraction by $<0.5\%$ where
non-zero.

\subsection{Robustness to a controlled estimator perturbation}
\label{sec:reliable-robust}

The pilot integrator resolves \texttt{glossy} and
\texttt{delta-mediated} through NEE only, suppressing the
BSDF-sampled emitter hit by an acceptance rule that avoids double
counting without MIS (Appendix~\ref{sec:appendix}); the energy
accounting on those buckets is by construction half of a complete MIS
estimator. If the argmax were sensitive to this, the label would
inherit exactly the estimator dependence we fault $\hat\sigma^2$ for.
We therefore restore the missing half --- a power-heuristic MIS on
the emitter-hit branch of those two buckets, the two on which the
suppressed half carries substantial energy on narrow lobes
(\texttt{emitter-direct} and \texttt{specular-direct} admit no
competition at all: primary rays have no NEE branch, delta vertices
zero NEE probability), leaving \texttt{direct}
and \texttt{diffuse-indirect} untouched so that the perturbation is
not conflated with the boundary flicker already characterised ---
and recompute the label at SPP$=4096$. The scope is deliberately
narrow: this is robustness to \emph{one} controlled perturbation,
the largest our own pilot construction introduces, not a
demonstration of general estimator independence; the argmax can in
principle still depend on sampling strategy, MIS heuristic, lobe
probabilities and omitted path families.

\begin{table}[!htbp]
\centering
\small
\caption{Robustness to estimator completion: per-pixel argmax
agreement between the NEE-only and MIS-completed estimators at
SPP$=4096$, and the 64-vs-4096 agreement recomputed under MIS.}
\label{tab:robust-mis}
\begin{tabular}{@{}lcc@{}}
\hline
scene & MIS agree (4096) & 64 vs 4096 (MIS) \\
\hline
\texttt{cubes\_dof}          & $100.00\%$ & $99.58\%$ \\
\texttt{tabletop\_separated} & $100.00\%$ & $99.42\%$ \\
\texttt{tabletop\_mixed}     & $100.00\%$ & $98.64\%$ \\
\texttt{glass\_caustic}      & $100.00\%$ & $95.64\%$ \\
\texttt{ring\_caustic}       & $100.00\%$ & $87.07\%$ \\
\texttt{snooker}             & $99.86\%$  & $89.58\%$ \\
\texttt{kitchen\_counter}    & $98.52\%$  & $89.87\%$ \\
\hline
\end{tabular}
\end{table}

Within that scope the argmax is nearly insensitive
(Table~\ref{tab:robust-mis}): five of seven scenes are unchanged to
the pixel --- including \texttt{ring\_caustic}'s $6{,}949$-pixel
\texttt{delta-mediated} population --- \texttt{snooker} moves by
$0.14\%$ and \texttt{kitchen\_counter} by $1.48\%$; cross-budget
agreement recomputed under MIS lands within a whisker of the NEE-only
values. Reweighting a bucket whose membership is a $\delta$-predicate
cannot flip it, and every flip that occurs lies on the energy
boundaries the MIS actually reweights ($401$ of $403$
\texttt{diffuse-indirect} departures on \texttt{kitchen\_counter} go
to \texttt{glossy}).

\begin{figure}[h]
  \centering
  \includegraphics[width=0.73\columnwidth]{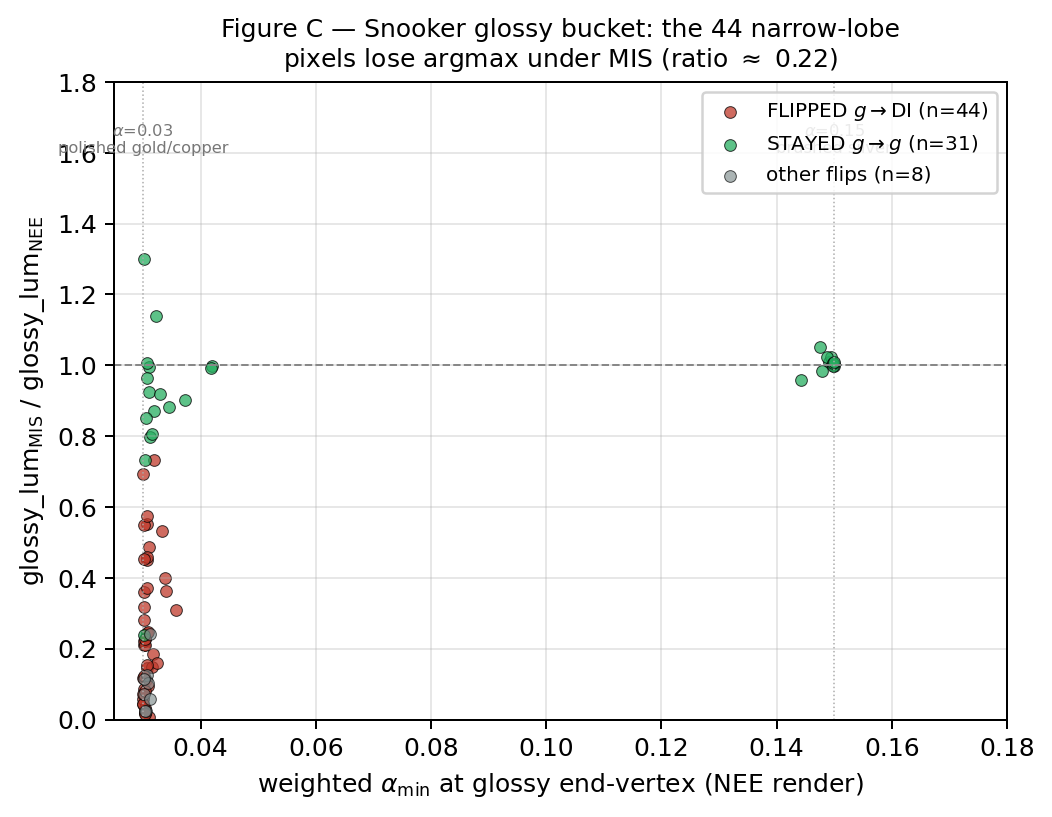}
  \caption{Mechanism of the \texttt{glossy} re-labelling on
    \texttt{snooker}. Each point is a pixel of the NEE-only
    \texttt{glossy} bucket, by end-vertex lobe width against the ratio
    of glossy energy under MIS to NEE-only. The $44$ flipped pixels
    (red) cluster at $\alpha\approx0.03$ (polished gold/copper), where
    MIS down-weights the NEE half and glossy energy contracts to
    ${\approx}0.22$ of its NEE-only value; the $31$ that stay (green)
    span wider lobes and keep a ratio near unity.}
  \label{fig:flip-mechanism}
\end{figure}

\paragraph{The flips are a correction, and they are lobe-width
dependent.}
On \texttt{snooker}, of the $83$ pixels the NEE-only pilot calls
\texttt{glossy}, $44$ re-label as \texttt{diffuse-indirect} under MIS
and $8$ move elsewhere, leaving $31$ (the MIS render's own
\texttt{glossy} bucket numbers $37$). The $44$ are the narrowest-lobe
pixels: on a near-specular lobe the BSDF PDF dominates the NEE PDF,
the power heuristic correctly drives the NEE weight toward zero, and
the NEE-only estimate --- which kept that half at full weight --- had
overstated glossy energy enough to push the argmax across the
boundary (Fig.~\ref{fig:flip-mechanism}). The lobe-width account
predicts that a glossy population not concentrated at the narrowest
lobes should be far more robust, and \texttt{kitchen\_counter} tests
this at scale: its $10{,}137$-pixel bucket, spanning
$\alpha\in[0.03,0.16]$, is $95.8\%$ stable under the same completion,
against $37\%$ for \texttt{snooker}'s $83$ pixels at
$\alpha\approx0.03$. The \texttt{snooker} fragility is the narrow-lobe
boundary behaviour the analysis says it is, not a property of the
glossy class; \S\ref{sec:reveal-lobe} isolates the variable.

\subsection{Stability on third-party scenes, predicted from the pilot}
\label{sec:reliable-transfer}

Every scene above is author-built. The third-party scenes address the
objection that the test distribution's generating process was under
our control. Predictions for each were fixed in writing before any
bucket-instrumented render of a third-party scene existed (one
amendment, registered after a first-batch outcome, is marked as
such in the supplementary material), and all are
reported against their outcomes, pass or fail, in the supplementary
material. Here we report the two whole-image results. Coverage
transfers: \texttt{other} receives exactly $0\%$ in every third-party
render. Stability transfers: 64-vs-4096 agreement is $0.978$ /
$0.990$ / $0.951$ on \texttt{veach-ajar} / \texttt{veach-mis} /
\texttt{bathroom2} (Table~\ref{tab:sentinel-stability}), inside the
first-party range.

\begin{figure*}[!t]
  \centering
  \includegraphics[width=0.92\textwidth]{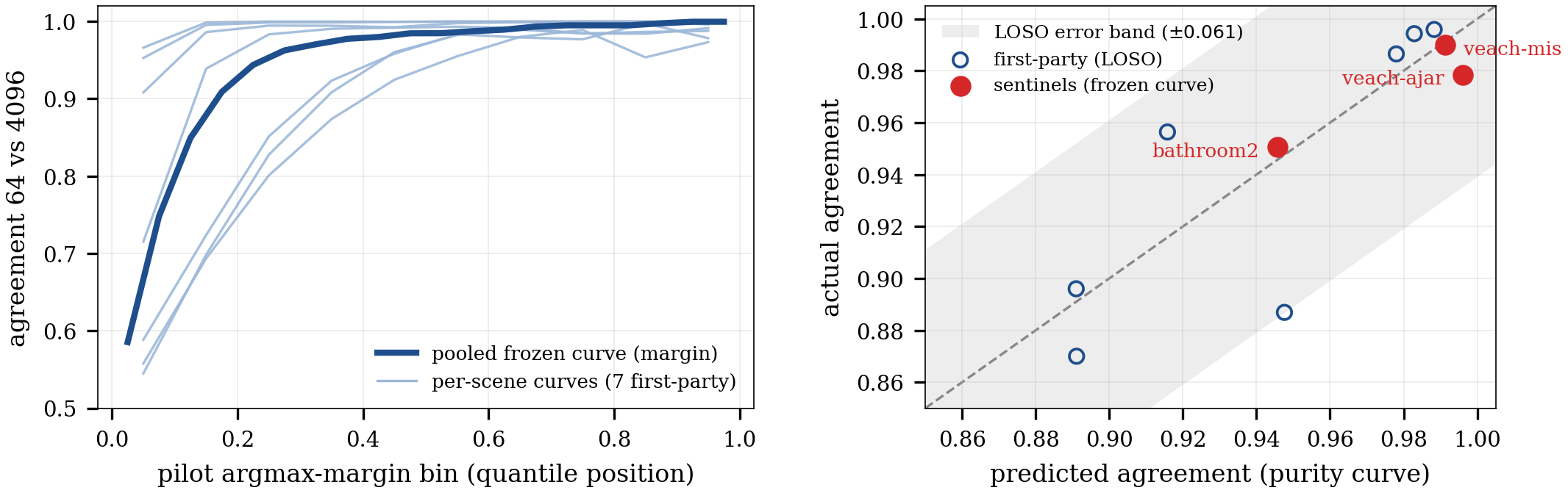}
  \caption{Pilot-based prediction of label stability. \emph{Left:}
    64-vs-4096 agreement as a function of the pixel's pilot
    argmax-margin bin, rising in every first-party scene (thin), and
    the pooled 20-bin monotone curve (bold), fitted on first-party
    data and registered before any third-party render. \emph{Right:} predicted versus actual
    scene-level agreement. Open circles: leave-one-scene-out
    predictions on the first-party scenes; the grey band is their
    observed maximum error ($\pm0.061$; the off-diagonal point is
    \texttt{kitchen\_counter}). Filled points: the three third-party
    scenes, each predicted from its own pilot --- errors
    $0.001$--$0.018$.}
  \label{fig:sentinel-predictor}
\end{figure*}

These values were predicted before the reference renders existed. On
the seven first-party scenes, agreement rises along pilot
argmax-margin (and purity) bins in every scene --- monotonically up
to a single late-bin fluctuation of $\le0.04$ on two --- and a pooled
20-bin monotone curve transfers across scenes with a
leave-one-scene-out scene-level error of at most $0.061$ (purity;
$0.079$ for margin), the worst case being the most mechanism-mixed
scene. The curve is fitted from first-party data alone and was
registered before any third-party render; each
scene's predicted agreement was issued from its own SPP-64 pilot
before its reference render existed, and
the outcomes lie within $0.018$ on \texttt{veach-ajar} and $0.005$ on
the other two (Fig.~\ref{fig:sentinel-predictor}). This suggests that
the label's stability on a new scene can be estimated from the
scene's own pilot before a reference exists. The error range is an
empirical one --- the maximum leave-one-scene-out error on
author-generated scenes --- and the three third-party outcomes are
consistent with it but do not constitute an independent re-estimate.

\begin{table}[!htbp]
\centering
\small
\caption{Third-party stability and frozen-curve predictions: 64-vs-4096
agreement predicted from each scene's own SPP-64 pilot via the pooled
margin/purity curves fitted on the first-party scenes.}
\label{tab:sentinel-stability}
\begin{tabular}{@{}lcccc@{}}
\hline
scene & agree & pred.\ (margin) & pred.\ (purity) & \texttt{other} \\
\hline
\texttt{veach-ajar} & $0.978$ & $0.995$ & $0.996$ & $0\%$ \\
\texttt{veach-mis}  & $0.990$ & $0.990$ & $0.991$ & $0\%$ \\
\texttt{bathroom2}  & $0.951$ & $0.950$ & $0.946$ & $0\%$ \\
\hline
\end{tabular}
\end{table}

% -------------------------------------------------------------------------
\section{Variance-Based Difficulty Read per Mechanism}
\label{sec:reveal}

We now use the descriptor to resolve the variance-based signals per
mechanism. Each subsection reports one observation that a scalar
per-pixel statistic does not expose.

\subsection{Variance reliability is tail-dependent and localises to
            mechanism}
\label{sec:reveal-reliability}

We measure two split-half quantities. First, a per-pixel convergence
exponent $\hat\alpha(p)$ is fit from $\hat\sigma^2$ at SPP
$\in\{64,256,1024,4096\}$ with ten independent seeds per level
($\sim$40{,}000 samples per pixel); the seeds are partitioned into two
halves, $\hat\alpha$ refit on each, and the two fields
Spearman-correlated, averaged over 20 random $5{+}5$ partitions
(partition-to-partition standard deviation $\le0.003$). Second, the
same procedure is applied to $\hat\sigma^2$ itself at each fixed SPP.
A noiseless reference would score $1.0$.

\begin{table}[!htbp]
\centering
\small
\caption{Split-half self-consistency of $\hat\sigma^2$: Spearman
$\rho$ between the fields of two disjoint five-seed halves, per fixed
SPP, mean over 20 partitions. Reliability is \emph{tail-dependent}: it
collapses on the caustic-dominated scene and is stable on
light-tailed ones.}
\label{tab:sigma-splithalf}
\begin{tabular}{@{}lcccc@{}}
\hline
scene & SPP=64 & 256 & 1024 & 4096 \\
\hline
\texttt{glass\_caustic} & 0.181 & 0.143 & 0.206 & 0.257 \\
\texttt{cubes\_dof} & 0.311 & 0.310 & 0.308 & 0.309 \\
\texttt{ring\_caustic} & 0.385 & 0.426 & 0.498 & 0.543 \\
\texttt{snooker} & 0.483 & 0.517 & 0.549 & 0.577 \\
\texttt{tabletop\_mixed} & 0.512 & 0.546 & 0.599 & 0.669 \\
\texttt{kitchen\_counter} & 0.553 & 0.601 & 0.619 & 0.630 \\
\texttt{tabletop\_separated} & 0.689 & 0.698 & 0.707 & 0.712 \\
\hline
\end{tabular}
\end{table}

The $\hat\alpha$-derived ranking on \texttt{cubes\_dof} and
\texttt{glass\_caustic} has split-half $\rho=0.23$--$0.29$ --- the
reliability figure of \S\ref{sec:intro}. The $\hat\sigma^2$
measurement (Table~\ref{tab:sigma-splithalf}) is sharper: its
self-consistency is a function of the scene's tail. On
\texttt{glass\_caustic} it sits at $0.18$--$0.26$, non-monotone in
SPP --- more samples do not buy stability, itself a heavy-tail
signature --- and the complementary cross-budget measurement on the
same scene, rank-correlating $\hat\sigma^2$ between the SPP-64 and
SPP-4096 renders, gives $\rho=0.26$. On light-tailed
\texttt{tabletop\_separated} it sits at $0.69$--$0.71$; on Lambertian
\texttt{cubes\_dof} a flat ${\sim}0.31$. Where the integrand is
light-tailed, $\hat\sigma^2$ is serviceable and nobody should stop
using it; where it is heavy-tailed --- the regime holding the hardest
pixels --- it fails to agree with itself across a seed split, and
evaluations against a $\hat\sigma^2$- or $\hat\alpha$-derived target
are attenuated accordingly. On the same scenes the mechanism label
agrees with its own reference at $87$--$99.6\%$ from a single 64-SPP
pilot.

\begin{table}[!htbp]
\centering
\small
\caption{Split-half self-consistency of $\hat\sigma^2$ at SPP$=4096$,
resolved \emph{per mechanism bucket} (mean over 20 partitions; ``---''
marks buckets with fewer than 50 pixels).}
\label{tab:sigma-perbucket}
\begin{tabular}{@{}lccccc@{}}
\hline
scene & dir. & diff. & d-m & spec. & gloss. \\
\hline
\texttt{snooker} & 0.50 & 0.65 & 0.41 & 0.59 & 0.83 \\
\texttt{glass\_caustic} & 0.18 & 0.42 & 0.13 & 0.46 & --- \\
\texttt{tabletop\_mixed} & 0.67 & 0.59 & 0.54 & 0.50 & --- \\
\texttt{tabletop\_separated} & 0.71 & 0.57 & 0.50 & 0.46 & --- \\
\texttt{cubes\_dof} & 0.20 & 0.72 & --- & --- & --- \\
\texttt{ring\_caustic} & 0.46 & 0.24 & 0.72 & 0.39 & --- \\
\texttt{kitchen\_counter} & 0.59 & 0.57 & 0.38 & 0.40 & 0.44 \\
\hline
\end{tabular}
\end{table}

\paragraph{Resolving it per bucket.}
The scene-level statement localises to the pixel level when the same
measurement is resolved per mechanism (Table~\ref{tab:sigma-perbucket}).
The clearest case is the \texttt{delta-mediated} bucket
of \texttt{glass\_caustic}: split-half $\rho=0.13$ at SPP$=4096$ and
$0.03$ at SPP$=64$ --- on the pixels the label marks as
delta-mediated in a caustic-dominated scene, the variance estimator
does not agree with itself across a seed split at all, and the pixels
were isolated into that bucket without ever looking at
$\hat\sigma^2$. The bucket is not uniformly the least reliable: in
the tabletops \texttt{specular-direct} is comparable, and on
\texttt{ring\_caustic} the bright focused cardioid makes it the
scene's \emph{most} reliable bucket at high SPP ($0.72$, from $0.32$
at SPP$=64$). The pattern is that within-bucket reliability tracks how
heavy-tailed the delta-mediated energy is at its pixels --- collapsing
where it is dim and firefly-borne, recovering where it is bright and
focused, and in between ($0.38$--$0.41$) where it is prominent but
mixed --- with the residual variation carried by the $\delta$-share
and luminance sidefields. A variance-based reference has no
representation to flag, in advance, which pixels carry this risk; the
\texttt{delta-mediated} bucket is a pre-computed map of where the
variance baseline is at risk, and the sidefields resolve where within
it the risk is realised.

One entry appears to contradict the cost analysis to follow:
\texttt{snooker}'s \texttt{glossy} bucket has the \emph{highest}
self-consistency ($0.83$) yet is slowest-converging
(\S\ref{sec:reveal-cost}). There is no contradiction: self-consistency
asks whether the \emph{estimator} agrees with itself, time-to-quality
whether the \emph{pixel} converges. A glossy pixel can carry a
stably-estimated, consistently high variance --- $\hat\sigma^2$
reliably reports ``hard'' --- whereas the delta-mediated failure is
that $\hat\sigma^2$ cannot report a consistent value at all. The
partition separates these two failure modes into two buckets where
one continuous number conflates them. (The $0.83$ is itself an average
over sub-populations: \S\ref{sec:reveal-lobe} shows the narrowest-lobe
glossy pixels have \emph{negative} $\rho$.)

\subsection{Mechanism is not difficulty: cost heterogeneity and
            censoring}
\label{sec:reveal-cost}

A mechanism class describes how energy arrives, not how expensive it
is to resolve, and per-bucket resolution makes the distinction
measurable. For
each pixel we compute a time-to-quality cost $c(p)$: the smallest SPP
at which $|\mathrm{lum}(I_N)-\mathrm{lum}(I_{\mathrm{ref}})|/
\max(\mathrm{lum}(I_{\mathrm{ref}}),\epsilon)<\tau$
($\tau=0.10$, $\epsilon=10^{-4}$, doubling SPP grid to $2048$,
independent renders per level, SPP-4096 reference). Pixels that fail
to cross by $2048$ are right-censored: their cost is unknown but above
the cap. Table~\ref{tab:cost-by-bucket} gives the median per bucket.

\begin{table}[!htbp]
\centering
\small
\setlength{\tabcolsep}{3.5pt}
\caption{Median time-to-quality cost (SPP) per dominant-label bucket.
``---'': no pixels. $^\dagger$: bucket heavily right-censored at
SPP$_{\max}=2048$ (rate below); a censored median is a lower bound,
not a measurement. The \texttt{glossy} population is NEE-defined
(\S\ref{sec:reliable-robust}).}
\label{tab:cost-by-bucket}
\begin{tabular}{@{}lccccccc@{}}
\hline
bucket & sn & gc & tm & ts & cb & rc & kc \\
\hline
\texttt{direct} & 16 & 16 & 16 & 16 & 16 & 32 & 32 \\
\texttt{diffuse-indirect} & 32 & 16 & 32 & 32 & 64 & 32 & 64 \\
\texttt{delta-mediated} & 32 & 32 & 32 & 32 & --- & 32 & 64 \\
\texttt{specular-direct} & 128 & 256 & 256 & 128 & --- & $2048^\dagger$ & $2048^\dagger$ \\
\texttt{glossy} & $2048^\dagger$ & --- & --- & --- & --- & --- & 128 \\
\hline
\multicolumn{8}{l}{\scriptsize $^\dagger$ censored: \texttt{glossy} on sn $51.8\%$;
\texttt{specular-direct} on rc $53.0\%$, on kc $48.8\%$.}
\end{tabular}
\end{table}

Three observations follow that a scalar signal does not expose. First, \textbf{the caustic is not the worst case}: on every
scene where it exists, \texttt{delta-mediated} converges at a median
of $32$--$64$~SPP --- including \texttt{ring\_caustic}, whose bright
cardioid it contains. The right-censored long tails live elsewhere:
in \texttt{snooker}'s narrow-lobe \texttt{glossy} bucket ($51.8\%$
unconverged at $2048$) and in the deep delta chains of
\texttt{specular-direct} on \texttt{ring\_caustic} ($53.0\%$, ring
interiors) and \texttt{kitchen\_counter} ($48.8\%$, mirror paths
among polished metals). Second, \textbf{the cost measurement itself
saturates} on exactly those buckets --- censoring means the
experiment has ceased to resolve costs above the cap, not that the
pixels have become constant in cost --- while the discrete label
remains discriminating on the same pixels. Third, \textbf{buckets are
internally heterogeneous}: \texttt{kitchen\_counter}'s
population-scale \texttt{glossy} bucket, spread across lobe widths,
converges at a median of $128$~SPP with only $5.9\%$ censoring,
whereas \texttt{snooker}'s $83$-pixel bucket, concentrated at
$\alpha\approx0.03$ and NEE-defined, is half censored; part of that
apparent hardness is the high variance of the NEE estimator on narrow
lobes rather than an intrinsic property of the transport --- itself an
instance of estimator dependence. A mechanism class is therefore not
an ordering of difficulty, and we do not claim one; it is a
coordinate system in which the heterogeneity becomes visible and
attributable. Per-dimension correlations between cost and the
continuous sidefields, reported in the supplementary material,
confirm the point: no single component is a stable cost proxy, and
the weak correlations are themselves a consequence of censoring on
the hardest pixels.

\subsection{Lobe width governs the glossy tail}
\label{sec:reveal-lobe}

\begin{figure}[h]
  \centering
  \includegraphics[width=0.75\columnwidth]{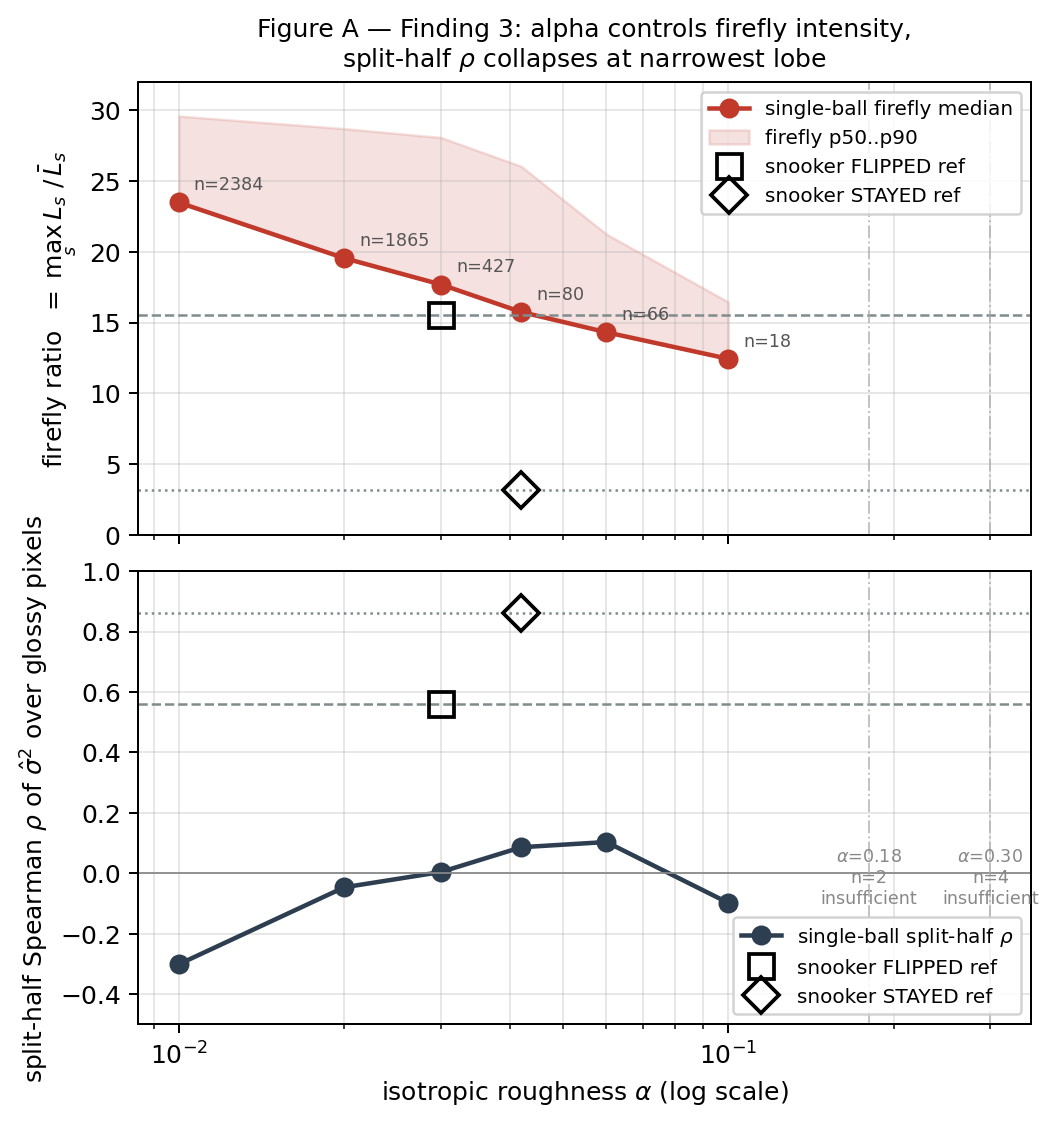}
  \caption{Controlled roughness sweep. \emph{Top:} median firefly
    ratio falls monotonically as the lobe widens, $23.5$ at
    $\alpha{=}0.01$ to $12.4$ at $\alpha{=}0.10$ (p50--p90 band); the
    \texttt{snooker} flipped pixels ($\square$, narrow lobe) land on
    the curve, the surviving pixels ($\Diamond$, wide lobe,
    multi-path) sit far below it. \emph{Bottom:} split-half $\rho$ of
    $\hat\sigma^2$ is not monotone in $\alpha$: it collapses to
    $-0.30$ at the narrowest lobe and otherwise sits in a noise band,
    while the surviving pixels' $\rho\approx0.86$ floats above the
    whole curve. The two widest levels yield too few
    \texttt{glossy}-dominant pixels to measure.}
  \label{fig:roughness-sweep}
\end{figure}

Both the estimator-perturbation flips (\S\ref{sec:reliable-robust})
and the cost censoring point at end-vertex roughness $\alpha$ as the
controlling variable of the \texttt{glossy} tail. A controlled
single-sphere experiment isolates it
(Fig.~\ref{fig:roughness-sweep}): a single isotropic
\texttt{roughconductor} sphere in an otherwise diffuse box --- no
delta surfaces, so no $\delta$-S contaminates the bucket --- with
$\alpha$ swept over eight values ($0.01$ to $0.30$; 30 seeds
$\times$ 128 SPP per level) and geometry, illumination and camera
fixed. The prediction, fixed before rendering, was that the per-pixel
firefly ratio (seed-wise maximum over mean glossy contribution;
statistics over glossy-dominant pixels, those whose dominant label
is \texttt{glossy}) and
the split-half $\rho$ of $\hat\sigma^2$ would both co-vary with
$\alpha$.

The firefly prediction holds cleanly: the median firefly ratio falls
monotonically from $23.5$ at $\alpha{=}0.01$ to $12.4$ at
$\alpha{=}0.10$ across the six populated levels. Narrow lobes carry a
heavier per-pixel tail --- the signature the kurtosis bound predicts,
the narrowest lobe being the highest-kurtosis integrand. The
reliability prediction holds only in its weaker form: the split-half
$\rho$ of $\hat\sigma^2$ does not rise smoothly with $\alpha$, but at
$\alpha{=}0.01$ it is $-0.30$ --- the two seed halves rank pixels in
\emph{anti}-correlation, the fingerprint of an estimator dominated by
rare large samples --- and elsewhere sits in a $\pm0.10$ noise band. We
claim no monotone $\rho(\alpha)$; what the controlled bucket shows is
that $\hat\sigma^2$ fails to agree with itself at every $\alpha$ we
can measure and agrees in reverse on the narrowest lobe. This is the
glossy counterpart of the delta-mediated failure above. One regime
resists the lobe-width account: \texttt{snooker}'s $31$ surviving
glossy pixels combine a low firefly ratio (${\approx}3$) with high
$\rho$ (${\approx}0.86$), which neither the sphere sweep nor a
light-count sweep at fixed $\alpha$ reproduces (supplementary); we
record it as open rather than claim a mechanism we have not isolated.

% -------------------------------------------------------------------------
\section{Cross-Scene Structure of the Descriptor}
\label{sec:structure}

The continuous sidefields turn each pixel into a small vector ---
$\delta$-share $C$, glossy fraction $G$, depth $D$ and inverse purity
$P=1-\pi$ --- and the correlations among these components, computed
per scene and compared across scenes, expose geometric variables a
scalar variance has no representation for. Table~\ref{tab:correlation-matrix}
reports all six pairs on all seven scenes; two of the sign reversals
are controlled by an identifiable variable and are examined here, the
rest are recorded in the supplementary material.

\begin{table}[!htbp]
\centering
\small
\setlength{\tabcolsep}{2.6pt}
\caption{Per-pixel Spearman $\rho$ among the four continuous
components, all seven scenes (cb, gc, sn, tm, ts, rc, kc:
\texttt{cubes\_dof},
\texttt{glass\_caustic}, \texttt{snooker}, \texttt{tabletop\_mixed},
\texttt{tabletop\_separated}, \texttt{ring\_caustic},
\texttt{kitchen\_counter}). ``\emph{deg.}'': one component
identically zero. $\ddagger$: sign reversal controlled by an
identified variable (\S\ref{sec:structure-visibility},
\S\ref{sec:structure-depth}); $\dagger$: sign-stable; $\circ$:
recorded observation (supplementary). Every cell is reported.
Three depth--purity and three depth--glossy entries differ from the
PG-submitted version, which cyclically mis-assigned them across
scenes --- a transcription error, corrected here from the
regenerated pipeline (revision report, Corrections~1).}
\label{tab:correlation-matrix}
\begin{tabular}{@{}lccccccc@{}l}
\hline
pair & cb & gc & sn & tm & ts & rc & kc & note \\
\hline
$C \leftrightarrow P$ & \emph{deg.} & $+0.22$ & $-0.09$ & $-0.30$ & $-0.35$ & $-0.18$ & $+0.18$ & $\circ$ \\
$P \leftrightarrow G$ & \emph{deg.} & \emph{deg.} & $-0.19$ & $-0.25$ & $-0.19$ & \emph{deg.} & $+0.55$ & $\circ$ \\
$C \leftrightarrow G$ & \emph{deg.} & \emph{deg.} & $-0.12$ & $+0.44$ & $-0.27$ & \emph{deg.} & $-0.02$ & $\ddagger$ \\
$D \leftrightarrow C$ & \emph{deg.} & $+0.73$ & $+0.45$ & $+0.43$ & $-0.21$ & $+0.29$ & $+0.13$ & $\circ$ \\
$D \leftrightarrow G$ & \emph{deg.} & \emph{deg.} & $+0.42$ & $+0.23$ & $+0.20$ & \emph{deg.} & $+0.87$ & $\dagger$ \\
$D \leftrightarrow P$ & $+0.66$ & $+0.52$ & $-0.37$ & $+0.24$ & $+0.50$ & $-0.40$ & $+0.60$ & $\ddagger$ \\
\hline
\end{tabular}
\end{table}

\subsection{Delta-mediated/glossy coupling follows mutual visibility}
\label{sec:structure-visibility}

A contribution event lands in \texttt{delta-mediated} if its sequence
contains a $\delta$-S event; the same path may also pass through a
glossy interaction, making $\gamma$ non-zero on the same pixel. The
two fractions therefore co-occur when scene topology makes combined
paths common and compete when geometry forces a path to pick one, so
the sign of their per-pixel correlation should depend on the
object-space separation of the two material families and cross zero
along it. ``Separation'' must be a measured quantity; we declare two:
the minimum surface-to-surface distance between any delta-S and any
glossy object, and the \emph{mutual visibility fraction} --- the
fraction of unoccluded segments between surface points sampled on the
two families (64 points per surface). The second is the
mechanism-relevant one, since co-occurrence requires a path segment
connecting the families.

\begin{table}[!htbp]
\centering
\small
\caption{Per-pixel Spearman correlation between the $\delta$-share and
glossy fraction against the two declared separation measures. The
ordering follows mutual visibility on the tabletop-family scenes;
metric distance alone does not order them (\texttt{snooker}'s closest
pair is nearer than \texttt{tabletop\_mixed}'s). \texttt{kitchen\_counter}
is an out-of-family point: high visibility but large metric
separation, $\rho\approx0$. Last row: third-party pre-registered test
(measures computed and sign interval fixed before rendering).}
\label{tab:sign-reversal}
\begin{tabular}{@{}lccc@{}}
\hline
scene & min dist.\,(m) & mut.\ vis. & $\rho_{C\text{--}G}$ \\
\hline
\texttt{tabletop\_mixed} & 0.15 & 0.094 & $+0.441$ \\
\texttt{snooker} & 0.06 & 0.078 & $-0.121$ \\
\texttt{tabletop\_separated} & 0.78 & 0.000 & $-0.270$ \\
\hline
\texttt{kitchen\_counter} & 0.53 & 0.125 & $-0.017$ \\
\hline
\texttt{bathroom2} (3rd-party, pre-reg.) & 0.34 & 0.327 & $+0.251$ \\
\hline
\end{tabular}
\end{table}

\begin{figure*}[!t]
  \centering
  \includegraphics[width=0.8\textwidth]{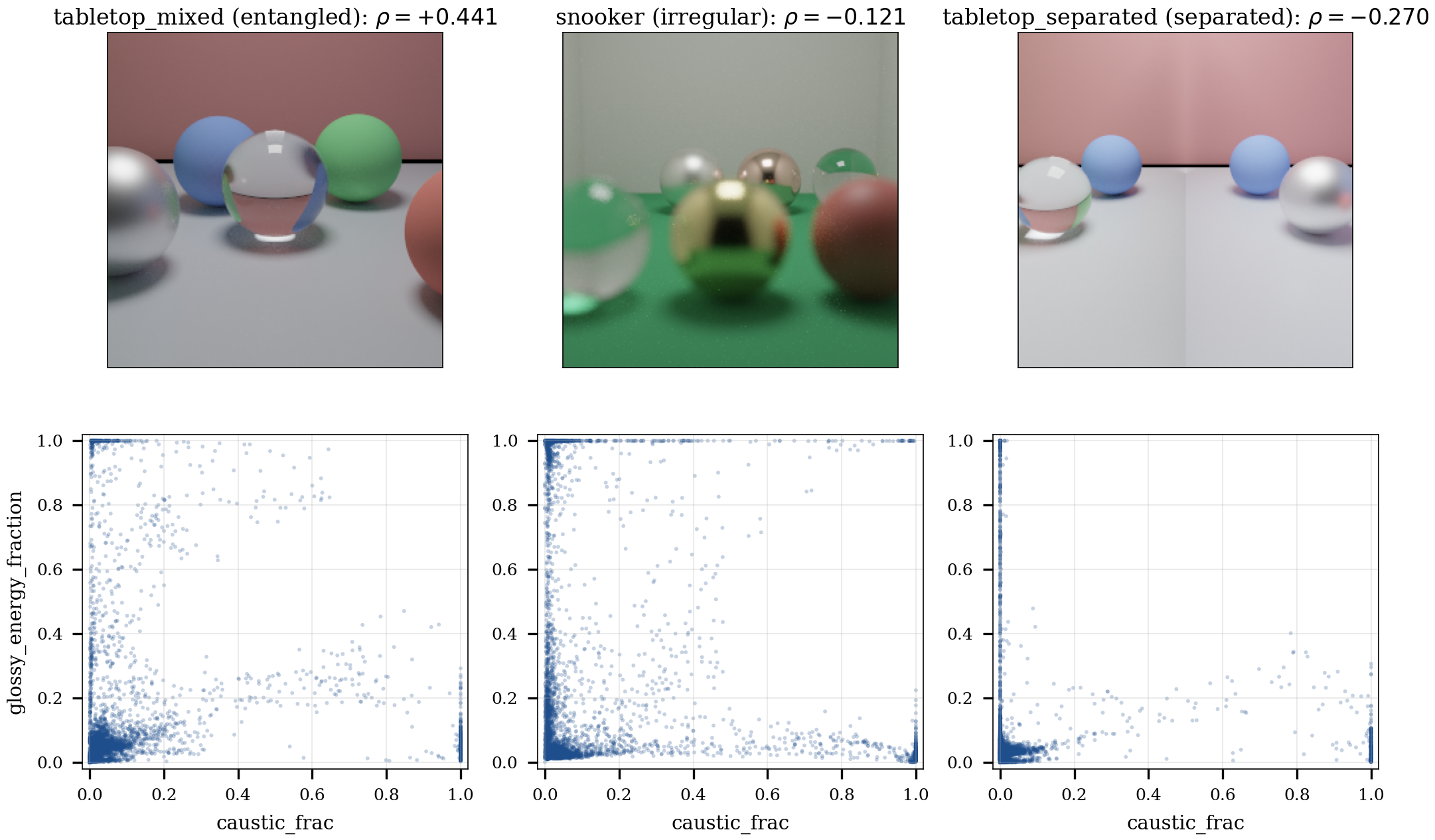}
  \caption{Sign reversal of the delta-mediated/glossy correlation with
    object-space separation. Top: the three scenes ordered by mutual
    visibility between the delta-S and glossy object families.
    Bottom: per-pixel scatter of $\delta$-share against glossy
    fraction; $\rho$ falls from $+0.441$ (entangled) through $-0.121$
    (\texttt{snooker}) to $-0.270$ (separated).
    \texttt{tabletop\_separated} is the controlled test: identical
    geometry and materials except for the sight-line-blocking
    partition.}
  \label{fig:sign-reversal}
\end{figure*}

The controlled pair carries the test: \texttt{tabletop\_separated}
shares \texttt{tabletop\_mixed}'s geometry and materials exactly and
adds an opaque partition that drops mutual visibility from $0.094$ to
$0.000$; the correlation reverses from $+0.441$ to $-0.270$, and
since the partition is the only change, the reversal on this pair is
causal (Table~\ref{tab:sign-reversal}, Fig.~\ref{fig:sign-reversal}).
Across the tested configurations the ordering follows mutual
visibility and crosses zero between \texttt{tabletop\_mixed} and
\texttt{snooker}. We state the claim at the strength the data
supports: an observed ordering with one controlled two-point contrast
--- not a continuous monotonic law, which would require a parametric
separation sweep. \texttt{kitchen\_counter} bounds the account: with
sight lines present but the families metrically distant, $\rho\approx0$
--- visibility is necessary for co-occurrence and coupling decays with
distance.

\paragraph{A pre-registered test on third-party geometry.}
For the \texttt{bathroom2} interior --- mirror and chrome fixtures
beside glossy tiles, an uncontrolled sight-line structure --- the two
separation measures were computed first (mutual visibility $0.327$,
the highest of any tested scene, at $0.34$~m), and the prediction
$0<\rho_{C\text{--}G}<+0.441$ --- positive by the visibility rule,
attenuated below the entangled anchor by the distance qualification
--- was registered before any bucket-instrumented render existed. The
outcome is $+0.251$: sign and interval as registered. An uninformed
sign guess succeeds half the time and the interval's upper end is the
largest value previously observed, so this is one successful
out-of-sample test, not a validated law. For the design question of
\S\ref{sec:descriptor-design} it suffices: the sign of the
relationship is a function of geometry that passes through zero, so
glossy stays its own bucket.

\subsection{Depth and purity: delta chains versus lobe mixing}
\label{sec:structure-depth}

If ``deeper paths are messier paths'', depth would predict mechanism
mixing and $\rho(D,P)$ would be positive on every scene. It is not:
$+0.66$, $+0.52$, $+0.24$, $+0.50$, $+0.60$ on \texttt{cubes\_dof},
\texttt{glass\_caustic}, \texttt{tabletop\_mixed},
\texttt{tabletop\_separated}, \texttt{kitchen\_counter}, but $-0.37$
on \texttt{snooker} and $-0.40$ on \texttt{ring\_caustic}. A single
mechanism accounts for the sign. Where deep paths are dominated by
$\delta$-S chains --- \texttt{snooker}'s inter-reflections among
chrome and glass balls, \texttt{ring\_caustic}'s multi-bounce
reflections inside the rings --- every added vertex is the same delta
event, so paths gain length without gaining lobe diversity: the
deepest pixels are the purest and $\rho(D,P)<0$. Where deep paths
accumulate through lobe-mixing interreflection --- diffuse bounce
chains, glass transmission into diffuse interiors, glossy--diffuse
mixing --- every added vertex can change lobe class, depth
accumulates mixture and $\rho(D,P)>0$. \texttt{ring\_caustic}, whose
only non-diffuse transport is delta chains, lands on the negative
side as the account requires. Depth and purity are independent
dimensions by data, not merely by construction; the account is
consistent with seven scenes but, except for \texttt{ring\_caustic},
derived from uncontrolled comparisons (\S\ref{sec:scope}).

% -------------------------------------------------------------------------
\section{Mechanism-Conditioned Pilot Correction for Sample Allocation}
\label{sec:buys}

The preceding sections establish that the descriptor is stable where
variance estimates are not, and that it localises where they fail.
This section asks whether that buys anything operational, with a
controlled equal-budget sample-allocation experiment whose
hypotheses, metrics and decision rules were fixed before the runs
(protocol summary, the full control set, and the companion budgets
in the supplementary material; the complete per-cell record,
including tail quantiles and allocation-stability diagnostics, in
the research artifact).

\subsection{Protocol and strategies}
\label{sec:buys-protocol}

Two-stage adaptive rendering at $256^2$ on all seven first-party
scenes. A pilot of $N_0\in\{4,16,64\}$ SPP yields both a pilot
variance $\hat\sigma^2(p)$ (luminance, unbiased) and pilot bucket
energies from which the pilot label $\hat b(p)$ is read --- label
estimation shares the pilot budget. Each strategy then allocates a
fixed extra budget of $32$~SPP average per pixel, Neyman-proportionally
($n(p)\propto\sqrt{D(p)}$, per-pixel baseline of $1$, identical totals)
from a strategy-specific score $D(p)$; the final image is the
per-pixel mean of the first $n(p)$ frames of a shared 256-frame pool
disjoint from every pilot, and error is measured against an
independent 4096-SPP reference. Every number is the paired outcome
over $K{=}10$ disjoint pilot draws: single-pilot comparisons can
reverse sign under re-piloting, so we treat multi-pilot pairing as
mandatory.

Baselines are \emph{uniform}, \emph{observed-var} ($D=\hat\sigma^2$,
the standard pilot-variance allocator) and an \emph{oracle} using the
pool's own converged variance. The proposed treatment is
\textbf{bucket-floor}: for pixels whose pilot label lies in the
heavy-tailed family \{\texttt{delta-mediated}, \texttt{specular-direct},
\texttt{glossy}\}, replace $\hat\sigma^2(p)$ by
$\max(\hat\sigma^2(p),\mathrm{med}_b)$, the median pilot variance
within the pixel's own bucket; light-tailed pixels are untouched. The
rationale is \S\ref{sec:reveal-reliability}'s: on heavy-tailed pixels
a small pilot usually \emph{misses} the tail, so $\hat\sigma^2$ is
biased low exactly where the label says the tail lives. Controls: a
\emph{random-partition placebo} (the same floor on a label map with
identical class sizes but shuffled assignment, fresh per draw) tests
whether coarse partitioning alone explains any gain; a
\emph{label-oracle} variant (labels from the reference) prices
pilot-label error. A \emph{robust-variance} incumbent,
$D=\hat\sigma^2_{\mathrm{MoM}}$ --- median-of-means variance over
$k=\lceil\sqrt{N_0}\rceil$ contiguous blocks of the pilot frames ---
and its floored variant \emph{robust-floor}, with its own matched
placebo, test whether robust estimation absorbs the effect; MoM is not
meaningfully defined at $N_0{=}4$ (two-sample blocks), a limitation
with content of its own, since at the smallest pilots mechanism
conditioning is the only correction that still applies. The label's
overhead is negligible: bucket-energy capture shares the pilot's path
tracing, the floor is a per-bucket median and a maximum ($<1$~ms per
image).

\begin{table}[!htbp]
\centering
\small
\setlength{\tabcolsep}{3.2pt}
\caption{Equal-budget allocation at pilot $N_0{=}4$: PSNR (dB, mean
over 10 disjoint pilot draws; seed-to-seed std $0.08$--$3.0$) and the
per-draw win rate of bucket-floor against observed-var on paired MSE
and paired p99 tail error. Bold: scenes on which the pilot-variance
allocator is at or below \emph{uniform}. On \texttt{cubes\_dof} (no
heavy-tailed buckets) bucket-floor reduces to observed-var exactly.}
\label{tab:downstream}
\begin{tabular}{@{}lcccccc@{}}
\hline
scene & uniform & obs-var & b-floor & oracle & MSE & p99 \\
\hline
\texttt{glass\_c.} & $\mathbf{30.79}$ & $30.01$ & $30.29$ & $36.64$ & 10/10 & 8/10 \\
\texttt{kitchen}   & $\mathbf{33.71}$ & $32.75$ & $33.03$ & $41.18$ & 9/10 & 10/10 \\
\texttt{snooker}   & $34.97$ & $37.83$ & $38.17$ & $42.46$ & 10/10 & 10/10 \\
\texttt{ring\_c.}  & $\mathbf{58.30}$ & $57.17$ & $57.50$ & $67.09$ & 7/10 & 8/10 \\
\texttt{t.\_mixed} & $37.57$ & $37.67$ & $38.08$ & $44.39$ & 10/10 & 2/10 \\
\texttt{t.\_sep.}  & $\mathbf{38.04}$ & $38.00$ & $38.35$ & $40.85$ & 10/10 & 7/10 \\
\texttt{cubes}     & $\mathbf{29.51}$ & $27.29$ & $27.29$ & $29.71$ & --- & --- \\
\hline
\end{tabular}
\end{table}

\begin{figure*}[!t]
  \centering
  \includegraphics[width=0.92\textwidth]{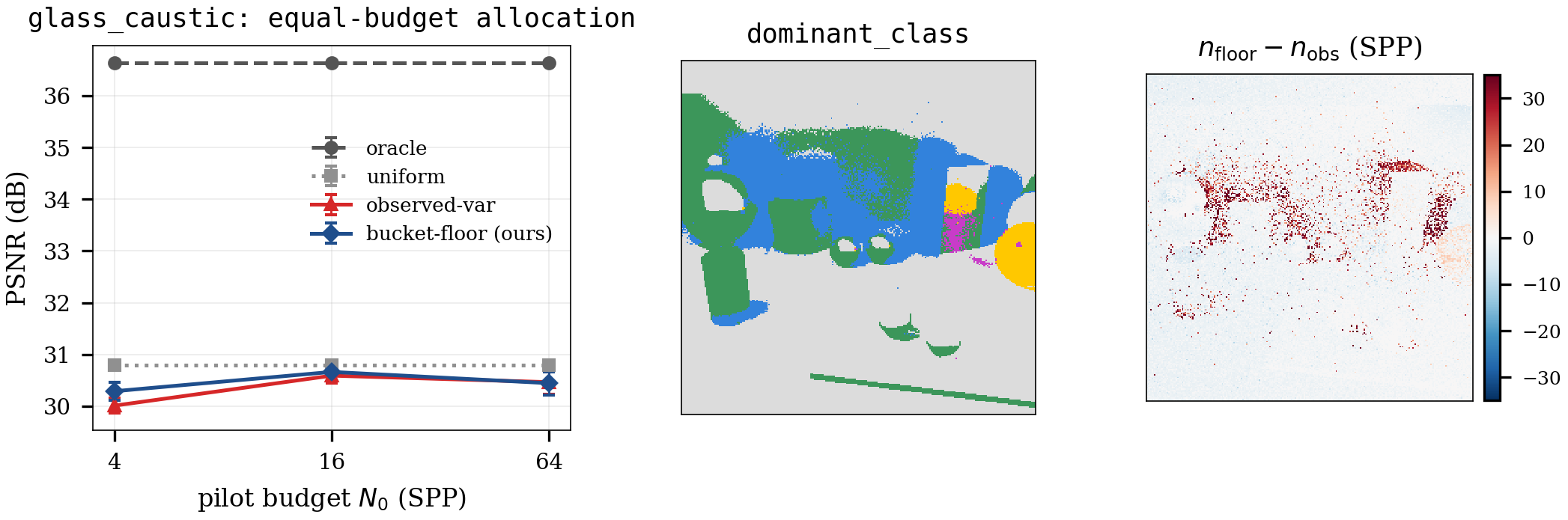}
  \caption{\emph{Left:} equal-budget PSNR versus pilot budget on
    \texttt{glass\_caustic} (mean $\pm$ std over 10 pilots). The
    pilot-variance allocator (red) sits below uniform (dotted) at every
    pilot budget, while the oracle (dashed) shows $+5.9$~dB of
    headroom; bucket-floor (blue) recovers part of the gap at small
    pilots and converges to the baseline as the pilot improves.
    \emph{Middle/right:} \texttt{kitchen\_counter} at $N_0{=}4$: the
    allocation difference $n_{\mathrm{floor}}-n_{\mathrm{obs}}$
    (right; red = more samples under bucket-floor) concentrates on the
    pixels the label marks \texttt{glossy} and \texttt{delta-mediated}
    (middle) --- the pixels whose pilot variance is systematically
    underestimated.}
  \label{fig:downstream}
\end{figure*}

\subsection{The pilot fails for two distinguishable reasons}
\label{sec:buys-fails}

At $N_0{=}4$ pilot-variance allocation sits at or below uniform on
five of seven scenes (Table~\ref{tab:downstream}), and the failures
decompose into two components. The first is \emph{generic
weight-estimation noise}: Neyman weights from four samples are noisy
on any scene, and where the true difficulty field is nearly
homogeneous there is little to gain from concentration and much to
lose from misplacing it. Lambertian \texttt{cubes\_dof} is the pure
case --- observed-var loses $2.22$~dB to uniform with no heavy-tailed
transport in the scene --- and \texttt{tabletop\_separated}'s
$-0.04$~dB is within seed noise. This component shrinks as the pilot
grows (\texttt{cubes\_dof} is within $0.15$~dB of uniform by
$N_0{=}64$), is repaired almost entirely by a per-pixel sample floor
of $4$ (\texttt{ring\_caustic}@4: $-1.13\to+0.02$~dB;
\texttt{cubes\_dof}@4: $-2.22\to-0.94$~dB; supplementary), and is
not what the descriptor addresses. The second is the \emph{heavy-tail
pathology} of \S\ref{sec:intro}: the pilot systematically
underestimates variance on the tail pixels. Its budget-persistent form
--- a loss to uniform that does not vanish as the pilot grows --- is
exhibited by \texttt{glass\_caustic} ($-0.20$~dB at $N_0{=}16$,
$-0.32$ at $64$; narrowing to $-0.04$ at a $4\times$ allocation
budget, so the persistence is strongest at the pilot and budget
scales adaptive sampling typically uses), the $n{=}1$ demonstration
it is; the other two heavy-transport deficits flip positive by
$N_0{=}16$. The tail-underestimation component is evidenced more
directly by bucket-floor's paired gains over observed-var (next
subsection) and by the oracle's $+5.9$~dB of unrealised headroom on
\texttt{glass\_caustic}. The deficits \emph{amplify} with the budget
being allocated ($-0.97\to-4.39$~dB on \texttt{kitchen\_counter}@4 as
the extra budget grows from $32$ to $128$~SPP): a noisy pilot does
more damage the more samples it is trusted with. We measure no-harm
against observed-var rather than uniform because the experiment
isolates the label's \emph{marginal} contribution to the incumbent;
whether a system should fall back toward uniform when its pilot is
weak is an orthogonal control, and not a trivial one --- a bucket-free
global-shrinkage placebo, which is exactly such a fallback,
underperforms observed-var on all seven scenes at $N_0{=}4$.

\subsection{Conditioning repairs the tail component, with specificity}
\label{sec:buys-repair}

On paired MSE, bucket-floor improves on observed-var in $7$--$10$ of
$10$ pilot draws on every first-party scene containing heavy-tailed
buckets
($+0.28$ to $+0.41$~dB at $N_0{=}4$; delta-mediated-bucket MSE halved
on \texttt{snooker}), and reduces to observed-var exactly on
\texttt{cubes\_dof}. Win counts are paired descriptive statistics,
not significance tests; under the stricter pre-fixed criterion
($\ge8/10$ wins \emph{and} mean gain above one seed standard
deviation, taking the larger of the two strategies') the $N_0{=}4$ MSE wins on \texttt{glass\_caustic} and
\texttt{tabletop\_separated} qualify, \texttt{kitchen\_counter} and
\texttt{snooker} qualify on the tail metric ($10/10$ p99 each), and
\texttt{ring\_caustic}'s $7/10$ stays within seed noise at $N_0{=}4$
and reaches $10/10$ at $N_0{=}16$. The two metrics can disagree
informatively: on \texttt{tabletop\_mixed}@4 bucket-floor lowers
delta-mediated-bucket MSE by $33\%$ yet slightly raises image-wide p99
in $8/10$ draws --- the floor spreads budget across the bucket while
observed-var concentrates on the few extreme pixels that set the
global quantile whenever its pilot happens to catch them.

The spatial structure of the correction is the predicted one
(Fig.~\ref{fig:downstream}, right): extra samples flow to the pixels
labelled \texttt{glossy} and \texttt{delta-mediated}. The
random-partition placebo reproduces none of the gains --- its paired
deltas are several times smaller or negative, and in every qualifying
cell below half the treatment's --- so the gain is attributable to
\emph{which} pixels share a bucket, not to quantisation. The
label-oracle variant performs within noise of the pilot-labelled one
($38.15$ vs $38.17$~dB on \texttt{snooker}@4): estimating the label
from the same 4-SPP pilot costs essentially nothing, which prices the
finite-sample-label concern of \S\ref{sec:descriptor-event}
operationally. Gains shrink as the pilot grows --- by $N_0{=}64$ every
delta lies within $\pm0.06$~dB except \texttt{ring\_caustic}'s
$+0.24$ --- the repair targets pilot noise and largely vanishes when
the pilot no longer needs repairing.

\subsection{Interaction with a robust variance estimator}
\label{sec:buys-robust}

\begin{figure}[!htbp]
  \centering
  \includegraphics[width=\linewidth]{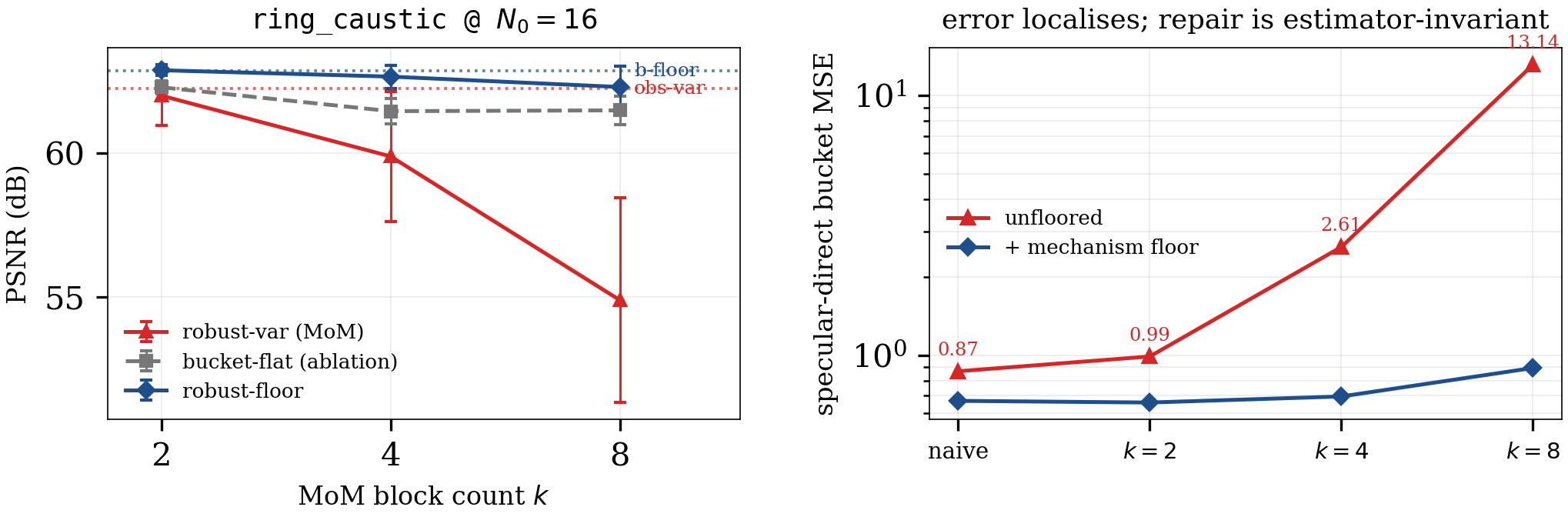}
  \caption{Robust estimation versus mechanism conditioning on
    \texttt{ring\_caustic} at $N_0{=}16$. \emph{Left:} PSNR versus
    median-of-means block count $k$: the robust incumbent (red)
    degrades monotonically as blocking becomes more aggressive; a
    per-bucket-uniform ablation (grey) recovers most of the loss ---
    the partition component --- and robust-floor (blue) holds a
    roughly constant level above it --- the per-pixel component.
    Dotted: the naive incumbent and its floored variant.
    \emph{Right:} the damage localises in \texttt{specular-direct},
    whose within-bucket ranking collapses (bucket MSE $0.99\to13.1$ as
    $k=2\to8$), while the floored level holds at $0.66$--$0.70$ for
    $k\le4$.}
  \label{fig:robust}
\end{figure}

The qualification of \S\ref{sec:related-variance} is here tested for
one robust estimator, median-of-means, and the result cuts both ways.
As an incumbent, $\hat\sigma^2_{\mathrm{MoM}}$ helps only where its
assumptions hold: it beats observed-var under the strict criterion in
exactly one cell (\texttt{glass\_caustic}@64, $9/10$), and on the most
delta-mediated scene it is harmful ($-2.35$~dB vs observed-var on
\texttt{ring\_caustic}@16): median-of-means gains stability by
discarding rare-tail evidence, and on single-firefly pixels that
evidence \emph{is} the signal. The mechanism floor is not absorbed by
the robust incumbent: robust-floor improves on robust-var on paired
MSE in $9$--$10$ of $10$ draws in eleven of twelve cells on the six
scenes with heavy-tailed buckets (Table~\ref{tab:robust-alloc}), and
its matched placebo reproduces none of it. A block-count sweep
localises the interaction (Fig.~\ref{fig:robust}): within-bucket rank
fidelity against the converged variance is flat in $k$ for
\texttt{delta-mediated} but collapses for \texttt{specular-direct}
(Spearman $0.52\to0.20$ on \texttt{ring\_caustic}@16 as $k=2\to8$),
the error explosion localises in that bucket ($0.99\to13.1$ bucket
MSE), and the mechanism-floored level is essentially
estimator-invariant ($0.66$--$0.70$ across the naive pilot and
$k\le4$). Choosing $k$ safely would require knowing in advance which
pixels are single-firefly dominated --- precisely what $\hat\sigma^2$
is least able to report about itself and what the label supplies for
free --- and no fixed $k$ serves both regimes, since mid-tail buckets
\emph{improve} under aggressive blocking (glossy rank fidelity
$0.33\to0.74$ on \texttt{snooker}@64) while rare-tail buckets degrade.
In the tested configuration, median-of-means estimation and
mechanism conditioning repair different failures, and the former
does not substitute for the latter on rare-tail pixels; other robust
constructions remain untested (\S\ref{sec:scope}).

\begin{table}[!htbp]
\centering
\small
\setlength{\tabcolsep}{3.4pt}
\caption{Pilot $N_0{=}16$: PSNR (dB, mean over 10 draws) for the
naive incumbent, the robust incumbent (MoM, $k{=}4$), and their
mechanism-floored variants; last columns: robust-floor's paired-MSE
win rate against robust-var at $N_0{=}16$ and $64$. Bold: the robust
incumbent falls below the naive one --- on heavy-tailed scenes
because blocking discards tail evidence, on \texttt{cubes\_dof}
because small-sample blocking is itself costly. On \texttt{cubes\_dof}
both floors are exact no-ops.}
\label{tab:robust-alloc}
\begin{tabular}{@{}lcccccc@{}}
\hline
scene & obs-var & MoM & b-floor & r-floor & @16 & @64 \\
\hline
\texttt{glass\_c.} & $30.59$ & $30.69$ & $30.66$ & $30.83$ & 10/10 & 6/10 \\
\texttt{kitchen}   & $34.14$ & $34.32$ & $34.34$ & $34.69$ & 9/10 & 10/10 \\
\texttt{snooker}   & $39.35$ & $\mathbf{39.31}$ & $39.57$ & $39.54$ & 10/10 & 9/10 \\
\texttt{ring\_c.}  & $62.24$ & $\mathbf{59.89}$ & $62.87$ & $62.66$ & 10/10 & 9/10 \\
\texttt{t.\_mixed} & $39.14$ & $\mathbf{38.72}$ & $39.17$ & $39.00$ & 10/10 & 9/10 \\
\texttt{t.\_sep.}  & $39.60$ & $\mathbf{39.29}$ & $39.67$ & $39.59$ & 10/10 & 10/10 \\
\texttt{cubes}     & $29.37$ & $\mathbf{28.79}$ & $29.37$ & $28.79$ & --- & --- \\
\hline
\end{tabular}
\end{table}

\subsection{Third-party geometry: \texttt{veach-ajar}}
\label{sec:buys-ajar}

The same protocol on the classical \texttt{veach-ajar} scene
(Table~\ref{tab:sentinel-ajar-alloc}) gives the largest pilot failure
measured in this work: observed-var loses to uniform by $6.8$~dB at
$N_0{=}4$ --- roughly three times the largest first-party deficit ---
with the oracle a further $10.5$~dB above observed-var. The fragility of
pilot-variance allocation is therefore not specific to our scene
construction. Bucket-floor ties observed-var ($-0.004$~dB, $5/10$),
for a reason the label itself reports: \texttt{veach-ajar} is
heavy-tailed in variance but light-tailed in mechanism --- its
difficulty lies in dim diffuse-indirect transport, and its
heavy-bucket population is ${\approx}1\%$ of pixels. The floor acts on
almost nothing, as designed. The label thus indicates, from the pilot
alone, that this failure belongs to the generic-starvation component
of \S\ref{sec:buys-fails} rather than the tail component it corrects;
a shared per-pixel sample floor, which targets starvation, recovers
$3.2$ of the $6.8$~dB (supplementary). The no-op behaviour observed on
\texttt{cubes\_dof} thus carries over to third-party geometry, and so
does the distinction the paper rests on: mechanism is not variance.

\begin{table}[!htbp]
\centering
\small
\caption{Equal-budget allocation on \texttt{veach-ajar} at $N_0{=}4$
(10 disjoint pilots). The pilot-variance allocator loses to uniform by
$6.8$~dB; bucket-floor ties observed-var because the scene's
heavy-bucket population is ${\approx}1\%$.}
\label{tab:sentinel-ajar-alloc}
\begin{tabular}{@{}lcc@{}}
\hline
strategy & PSNR (dB) & vs.\ obs-var \\
\hline
uniform          & $17.23 \pm 0.00$ & $+6.79$ \\
observed-var     & $10.45 \pm 0.46$ & --- \\
bucket-floor     & $10.44 \pm 0.46$ & $-0.004$ (5/10) \\
oracle           & $20.93 \pm 0.00$ & $+10.48$ \\
\hline
\end{tabular}
\end{table}

\subsection{What this demonstrates, and what it does not}
\label{sec:buys-scope}

No pilot-based strategy in Table~\ref{tab:downstream} --- ours
included --- beats uniform sampling on all seven scenes at $N_0{=}4$,
and on the heavy-transport scenes where the pilot fails outright,
bucket-floor narrows but does not close the gap to uniform. The
demonstrated claim is improvement over the pilot-variance allocator:
bucket-floor improves on observed-var wherever heavy-tailed buckets
carry appreciable population and reduces to it where they are absent,
so a system already running pilot-variance allocation loses nothing
in our tests and gains on the heavy-transport scenes. The oracle's
$+5.9$ to $+8.8$~dB headroom over uniform on those scenes locates the
remaining loss in pilot \emph{quality}, not in the
variance-proportional principle. The experiment shows the descriptor
in the role argued for: a stable conditioning variable that
identifies a priori the pixels on which a continuous estimate needs
repair --- not a replacement difficulty scalar and not a complete
allocator. It is one demonstration on one task with deliberately
simple estimators; richer uses (denoising auxiliaries, population-model
corrections in the spirit of \cite{sakai2025stater}, Bayesian
shrinkage with mechanism priors, per-bucket adaptive block counts)
remain open.

% -------------------------------------------------------------------------
\section{Scope and Limitations}
\label{sec:scope}

\paragraph{Taxonomy scope.}
The descriptor covers surface path tracing under a stated D/G/S
decomposition. It does not address participating media, subsurface
transport, hair and foliage visibility complexity, glints,
environment emitters or measured materials with non-unique lobe
decompositions --- several of which are major difficulty sources in
production; extending the taxonomy there, e.g.\ with phase-function
lobes as an additional axis, is future work. Coverage of the named
mechanisms is established on the tested scene class only.

\paragraph{What the label sees, and what the sidefields carry.}
Two third-party scenes test the taxonomy's boundary. On
\texttt{veach-mis} --- four metal plates of graded roughness under a
graded light row, the scene designed to stress NEE/BSDF balance --- the
plate faces are dominated by \texttt{diffuse-indirect} ($13{,}825$ of
$13{,}863$ pixels), not \texttt{glossy}, contrary to our prediction:
what a rough plate mostly shows the camera is a glossy-mediated
\emph{reflection of the diffuse surround}, camera $\to$ plate (G)
$\to$ diffuse end-vertex, which the end-vertex rule routes to the
diffuse end-vertex by design. Camera-side glossy mediation is
invisible to the dominant label by construction and carried entirely
by $\gamma$ (plate-face glossy fraction $0.94$--$1.00$ against
${\approx}0.03$ elsewhere). On \texttt{matpreview} with the
\texttt{principled} BSDF's clearcoat swept over $\{0,0.25,0.5,1.0\}$,
the NEE rule routes the coated material to \textbf{D} at every coat
value, agreement is $0.989$ throughout and the sweep moves $0.2\%$ of
pixels, while $\gamma$ responds monotonically (mean $0.040\to0.047$;
90th percentile $0.166\to0.196$). Both are design consequences, not
label instability: the discrete label holds the main axis still and
the continuous mixture carries the rest. Two of our sharpened
per-plate predictions on \texttt{veach-mis} failed for the same
reason --- they misread the taxonomy's scope --- and are reported as
registered in the supplementary material. Measured BRDFs remain
untested.

\paragraph{Convention and estimator dependence.}
The per-event label depends on renderer conventions; the per-pixel
label is a finite-sample estimator. We measured robustness to budget
changes and to one controlled estimator perturbation; bidirectional,
guided and regularised estimators are untested.

\paragraph{Findings.}
The visibility account (\S\ref{sec:structure-visibility}) rests on a
controlled pair, an ordering, and one blind prediction --- not a
parametric separation sweep. The depth/purity account is consistent
with seven scenes but, except for \texttt{ring\_caustic}, derived from
uncontrolled comparisons. Two further sign-varying pairs and the
stable-glossy sub-population of \texttt{snooker} remain open
(supplementary). By construction the \texttt{direct} bucket admits
both Lambertian single-bounce transport and single-bounce glossy
highlights; on \texttt{kitchen\_counter} a $\gamma$-split isolates a
$4.7\%$ glossy-lit sub-population with twice the median cost and half
the $\hat\sigma^2$ reliability of its Lambertian complement
(supplementary). An explicit lobe sub-split of \texttt{direct} would
resolve this without disturbing the main partition.

\paragraph{Downstream.}
The allocation experiment uses a simple NEE-only pilot, a two-stage
budget split and one robust incumbent (median-of-means, chosen for
having no free parameter beyond the block count); Catoni-type
$M$-estimators and production adaptive pipelines are untested, and we
deliberately do not evaluate the label as a denoiser auxiliary
channel, where any conclusion would be dominated by the choice of
denoiser family. The split-half methodology applies to denoiser
guidance maps directly and is the form in which we expect the
descriptor to be useful there first.

% -------------------------------------------------------------------------
\section{Conclusion}
\label{sec:conclusion}

Per-pixel rendering difficulty has had thirty-five years of
operational consensus around a single noisy scalar. We have argued
that difficulty should be characterised through transport structure,
with variance treated as a measurement whose reliability that
structure helps predict, and we have supported the argument with
measurements. The finite-sample variance is governed by the
integrand's fourth moment, of low split-half reliability as an
evaluation target, and least reliable on the pixels a mechanism
descriptor isolates in advance. The descriptor itself is stable
across a $64\times$ budget change where quantile-binned variance is
not, survives a controlled estimator perturbation, and on three
third-party scenes its stability was predicted from each scene's
pilot. Read per mechanism, difficulty shows structure a scalar hides:
in our scenes the caustic is not the most expensive transport, lobe
width governs the glossy firefly tail, and coupling between
mechanisms follows scene geometry. Conditioning on the descriptor has
measurable value: a pilot variance corrected by the label improves
equal-budget allocation wherever heavy-tailed buckets carry
appreciable population, reduces to the incumbent where they are
absent (exactly) or negligible (to within $0.004$~dB), and repairs
a failure that the
tested median-of-means estimator does not. Under heavy-tailed
transport, finite-sample variance is an unstable measurement; the
transport mechanism of each contribution --- deterministic under
stated renderer conventions --- is a stable structure to characterise
difficulty by, and conditioning on it yields a measurable gain.

Several extensions follow. A perceptual variant of the cost analysis
(FLIP, FovVideoVDP) may censor differently; a Bayesian shrinkage
formulation with mechanism as prior and sample statistics as
likelihood is the natural composition; a per-bucket adaptive
median-of-means block count is motivated directly by
\S\ref{sec:buys-robust}; and a richer explicit path-length axis would
complement the lobe-based labels along a dimension orthogonal to
mechanism type.

% -------------------------------------------------------------------------
\appendix
\section{Reproducibility}
\label{sec:appendix}

All first-party results were produced with Mitsuba~3.8.0
(\texttt{cuda\_ad\_rgb}) on a single NVIDIA RTX~5090; the third-party
suite ran under the same renderer version on an RTX~4090, recorded per
render. Four auxiliary measurements predate the uniform re-render
and are used unchanged from the original runs: the cross-budget
convergence study behind $\rho=0.26$, the $\hat\alpha$ reliability
estimate, the \texttt{snooker} glossy sub-population seed renders,
and the light-count control. The complete first-party pipeline re-runs in under one hour. This
appendix and the supplementary material give the implementation and
protocol specifications in detail; the corresponding code --- the
custom integrator, all procedural scene builders, analysis scripts,
seed layouts, and the allocation protocol, from whose entry points
every table and figure regenerates --- is being consolidated into a
public research artifact and will be released separately.

\paragraph{Classifier.}
Per-event classification follows Table~\ref{tab:partition} with the
sampled-lobe cascade Delta${}>{}$Diffuse${}>{}$Glossy for BSDF-sampled
bounces and the flag rule of \S\ref{sec:descriptor-event} for NEE
events. BSDF-to-lobe mapping: \texttt{diffuse}${}\to{}$D;
\texttt{roughplastic}${}\to{}$D under NEE, sampled lobe otherwise;
\texttt{roughconductor}, \texttt{roughdielectric}${}\to{}$G;
\texttt{conductor}, \texttt{dielectric} (smooth)${}\to{}$S;
\texttt{twosided} transparent to flags. Contribution energy is
Rec.~709 luminance of the full-throughput contribution (post
Russian-roulette compensation, post MIS weight where applicable);
zero-energy pixels carry no label; argmax ties break to the lower
bucket index (measure zero). Contributions are non-negative by
construction; non-finite values do not arise in the bucket
accumulators, and rare non-finite pixels in raw float32 RGB frames
are sanitised to zero by the analysis loaders and counted as
background. The \texttt{delta-mediated} bucket and the $\delta$-share
sidefield carry the keys \texttt{caustic} and \texttt{caustic\_frac}
in the implementation.

\paragraph{Estimator.}
NEE at every Smooth-flagged vertex (one NEE and one BSDF sample per
vertex); BSDF-into-emitter accepted iff the
primary ray or the previous bounce was delta; Russian roulette from
depth 5 at $p=\min(\max\beta,0.95)$; max depth 12; $256^2$, box
filter, independent sampler. Restricted MIS: power heuristic
($\beta{=}2$, solid-angle measure) on the NEE and emitter-hit branches
of the \texttt{glossy} and \texttt{delta-mediated} buckets only.

\paragraph{Statistics.}
$\hat\sigma^2$ at budget $K$ is $K\cdot\mathrm{Var}$ across 10 seeds
(ddof 1) of per-seed luminance; background dropped at mean luminance
$\le10^{-4}$; split-half values are means over 20 random $5{+}5$
partitions (base seed 0). Quantile bins: per scene, independently at
each SPP. Weighted $\kappa$ as in Table~\ref{tab:kappa}.
Time-to-quality: $\tau=0.10$, $\epsilon=10^{-4}$, SPP grid
$\{16,\dots,2048\}$ doubling, independent renders per level,
first-crossing rule, reference the seed-disjoint SPP-4096 stability
render. Convergence exponent $\hat\alpha(p)$: per-pixel unweighted OLS
of $\log_{10}$ luminance variance on $\log_{10}$ SPP over four levels,
$\hat\alpha=-$slope; pixels with non-positive or non-finite variance
at any level excluded; per-pixel $R^2$ is stored alongside. Separation measures: minimum surface distance
and mutual visibility over 64 surface points per object pair
(ray-tested, offset $10^{-3}$, seed 0); on \texttt{bathroom2} a
$20$k-pair unbiased subsample, family assignment by BSDF flags.
Roughness sweep: $\alpha\in\{0.01,0.02,0.03,0.042,0.06,0.10,0.18,0.30\}$,
30 seeds $\times$ 128 SPP per level plus a 4096-SPP reference at
$384^2$. Allocation protocol: seed layout pool${}={}$1000${}+{}i$,
pilot draw $k$ frame $j$ ${}={}$100000${}+{}1000k{+}j$, reference
chunks ${}={}$900000${}+{}c$, disjoint by construction; median-of-means
over $k=\lceil\sqrt{N_0}\rceil$ contiguous render-order blocks
(per-block variance ddof 1), sensitivity $k\in\{2,4,8\}$ at
$N_0{=}16$ and $\{4,8,16\}$ at $N_0{=}64$; pilot frames re-rendered
with per-frame luminance storage under identical seeds, SHA-256
manifests verifying every pool and pilot statistic bit-identical.
Third-party scenes: Bitterli's collection in native Mitsuba-3 XML,
$448\times256$, paper estimator configuration overriding native
defaults; \texttt{matpreview}'s environment map replaced by an
overhead $3{\times}3$ area panel at radiance $7.2$ (mean image
luminance $0.38$, matching the original). All predictions for the
third-party scenes, the frozen reliability curves, and every
prediction/outcome file are retained for inclusion in the public
artifact and summarised in the supplementary material.

% -------------------------------------------------------------------------

% -------------------------------------------------------------------------
\bibliographystyle{unsrt}
\bibliography{references}

% =========================================================================
% SUPPLEMENTARY MATERIAL
% (was supplementary.tex, a standalone CGF document; merged into this PDF)
% Numbering restarts as S1, S2, ... for sections, figures and tables.
% =========================================================================
\clearpage

\setcounter{section}{0}
\setcounter{figure}{0}
\setcounter{table}{0}
\renewcommand{\thesection}{S\arabic{section}}
\renewcommand{\thetable}{S\arabic{table}}
\renewcommand{\thefigure}{S\arabic{figure}}
% keep hyperref anchors unique now that the counters restart
\renewcommand{\theHsection}{S\arabic{section}}
\renewcommand{\theHtable}{S\arabic{table}}
\renewcommand{\theHfigure}{S\arabic{figure}}

\fancyhead[LO]{Supplementary Material --- Characterising Per-Pixel Rendering Difficulty by Transport Mechanism}

\begin{center}
  \rule{\textwidth}{2pt}
  \vskip 0.25in
  {\LARGE\sc Supplementary Material}\\[0.45em]
  {\large\sc for Characterising Per-Pixel Rendering Difficulty\\[0.2em]
   by Transport Mechanism}
  \vskip 0.29in
  \rule{\textwidth}{2pt}
\end{center}
\vskip 0.15in

\begin{quote}
This document collects material referenced from the main paper that
supports, but is not required for, its argument: additional
correlation observations and controlled sweeps
(\S\ref{sup:structure}), the per-dimension cost analysis
(\S\ref{sup:cost}), within-bucket heterogeneity of the \texttt{direct}
bucket (\S\ref{sup:direct}), per-bucket estimator-perturbation
robustness (\S\ref{sup:robust}), the full control set of the
allocation experiment (\S\ref{sup:alloc}), and the complete record
of the third-party scene suite, including every registered
prediction and its outcome (\S\ref{sup:sentinels}). Section numbers
of the form \S$n$ refer to the main paper; \S S$n$ to this document.
\end{quote}
\vskip 0.15in

% -------------------------------------------------------------------------
\section{Additional Correlation Observations and Sweeps}
\label{sup:structure}

\paragraph{Two further sign-varying pairs.}
Main-paper Table~9 records two pairs whose sign varies across scenes
without, so far, an identified controlling variable.
\emph{Depth versus $\delta$-share} runs $+0.73$, $+0.45$, $+0.43$,
$+0.29$, $+0.13$ across five non-degenerate scenes but $-0.21$ on
\texttt{tabletop\_separated}. The positive values are unsurprising ---
delta-mediated paths include a specular bounce and so tend to be at
least one vertex longer than direct paths. The negative outlier is
harder to interpret with confidence: the partition geometry
constrains both depth (paths must detour around the wall) and the
$\delta$-share (which surfaces the partition occludes)
simultaneously, and a single-scene anomaly does not distinguish a
genuine mechanism from a partition-induced artefact.
\emph{Inverse purity versus glossy fraction} is mildly negative on
\texttt{snooker} and the tabletops ($-0.19$ to $-0.25$) but $+0.55$
on \texttt{kitchen\_counter}, where glossy energy arrives layered
over direct and diffuse transport on the same pixels, so glossy-heavy
pixels are also mixed pixels. Both observations would need controlled
sweeps analogous to the tabletop pair to be promoted to findings.
The \emph{$\delta$-share versus inverse purity} pair is weak
($|\rho|\le0.35$) on every non-degenerate scene.

\paragraph{What lobe width does not explain.}
The \texttt{snooker} \texttt{glossy} bucket contains a
sub-population --- the $31$ pixels that survive MIS completion as
\texttt{glossy} (main paper \S4.4) --- with both a low firefly ratio
(${\approx}3$) and a high split-half $\rho$ (${\approx}0.86$), a
``stable'' glossy regime that the single-sphere roughness sweep (main
paper \S5.3) never reproduces at any $\alpha$. We tested whether
per-pixel path multiplicity was the missing variable by holding
$\alpha{=}0.03$ fixed and sweeping the number of energy-conserving
light sources from one to sixteen. It is not: the firefly ratio
stayed flat (as predicted, since $\alpha$ was fixed), but $\rho$ did
not climb out of the noise band, because adding light \emph{sources}
did not increase per-pixel supplying-path \emph{multiplicity} --- each
pixel still saw only one or two caustic hits. The stable-glossy
regime of \texttt{snooker} is thus governed by some scene-topology
variable --- multi-object inter-reflection is our leading candidate ---
that neither controlled scene reproduces; isolating it requires a
controlled two-object experiment.

% -------------------------------------------------------------------------
\section{Cost Versus the Descriptor Dimensions}
\label{sup:cost}

\begin{figure}[!htbp]
  \centering
  \includegraphics[width=0.8\columnwidth]{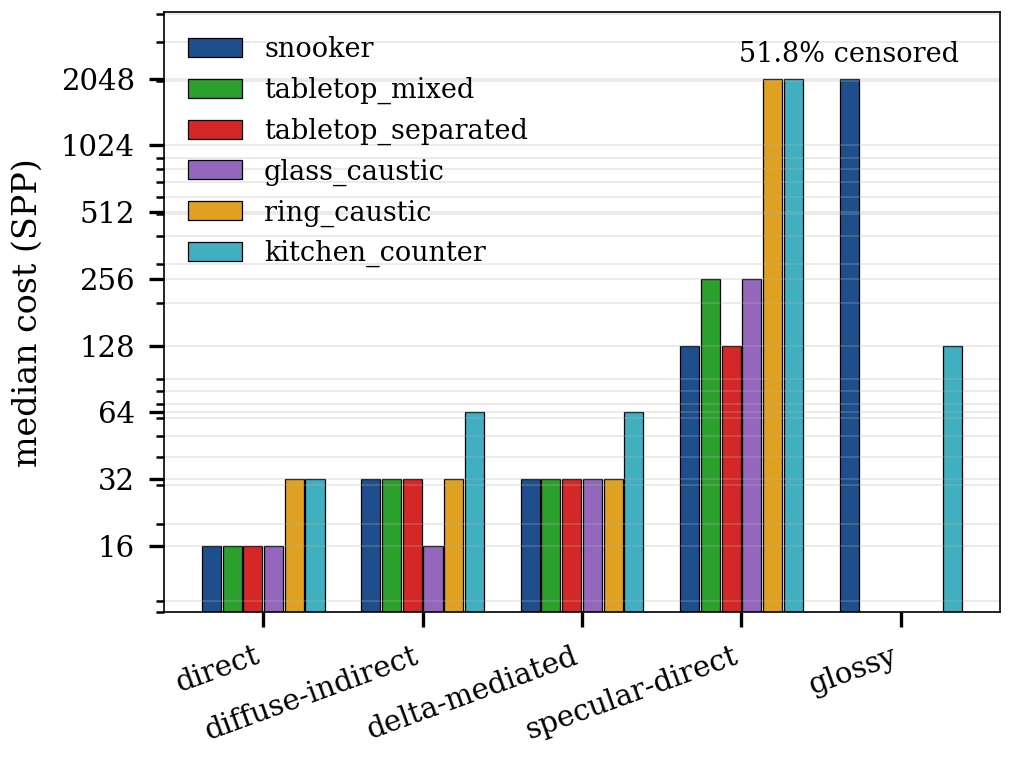}
  \caption{Median time-to-quality cost per dominant-label bucket
    across the six scenes with non-trivial transport (main-paper
    Table~8 as a figure). Three censored medians sit at the
    SPP$=2048$ cap: the NEE-defined \texttt{glossy} bucket on
    \texttt{snooker} ($51.8\%$ right-censored) and the deep-chain
    \texttt{specular-direct} buckets on \texttt{ring\_caustic}
    ($53.0\%$) and \texttt{kitchen\_counter} ($48.8\%$), while
    \texttt{kitchen\_counter}'s population-scale \texttt{glossy}
    bucket converges at a median of $128$~SPP.}
  \label{sfig:cost-by-bucket}
\end{figure}

Main-paper \S5.2 reports per-bucket time-to-quality cost
(per-bucket medians in Fig.~\ref{sfig:cost-by-bucket}).
Table~\ref{stab:cost-perdim} asks the complementary question: whether
the continuous sidefields correlate with cost in a stable way ---
whether depth, $\delta$-share, inverse purity or glossy fraction would
each serve, alone, as a cheap cost proxy. They would not, and the
reasons are informative about both the cost measurement and the
vector.

\begin{table}[!htbp]
\centering
\small
\caption{Per-pixel Spearman correlation between time-to-quality
$c(p)$ and the four continuous components, all seven scenes (column
abbreviations as in the main paper). ``n/a'': no energy in that
scene. Every cell is reported.}
\label{stab:cost-perdim}
\setlength{\tabcolsep}{2.2pt}
\begin{tabular}{@{}lccccccc@{}}
\hline
dim & sn & gc & tm & ts & cb & rc & kc \\
\hline
depth         & $+0.27$ & $+0.22$ & $+0.44$ & $\mathbf{+0.04}$ & $+0.35$ & $+0.22$ & $+0.34$ \\
$\delta$-share & $+0.07$ & $+0.17$ & $+0.21$ & $+0.01$ & n/a & $+0.15$ & $+0.05$ \\
inv.\ purity   & $-0.16$ & $+0.06$ & $+0.07$ & $-0.10$ & $+0.09$ & $+0.05$ & $+0.26$ \\
glossy frac.  & $+0.23$ & n/a & $+0.22$ & $+0.20$ & n/a & n/a & $+0.29$ \\
\hline
\end{tabular}
\end{table}

\paragraph{Glossy fraction is the only sign-stable dimension, and its
magnitude is structurally capped.}
$\rho(\text{glossy frac.},c)$ is positive on all four non-degenerate
scenes, in the narrow range $+0.20$ to $+0.29$; no other component is
as stable. But $\rho\approx0.2$ explains only ${\sim}4$--$8\%$ of the
variance in $c$, and on \texttt{snooker} this is not the intrinsic
strength of the relationship: with $51.8\%$ of the NEE-defined glossy
bucket right-censored, Spearman $\rho$ assigns those pixels the same
maximal rank, compressing the top of the distribution into a plateau
that pulls correlations toward zero. The weak $\rho$ does not mean the
descriptor fails to track cost; it means cost itself has gone flat on
the very pixels the vector flags as glossy-dominant.

\paragraph{Inverse purity versus cost is sign-unstable.}
$\rho(\text{inv.\ purity},c)$ runs $-0.16$ to $+0.26$ across the
seven scenes --- mostly small, signs disagreeing. The instability is
consistent within each scene but needs care: the link between
mechanism mixing and cost is mediated by which bucket the mixed pixels
fall into and which buckets are expensive there. The component flags
mixture; it is not, alone, a stable cost proxy.

\paragraph{Depth versus cost is geometry-driven.}
$\rho(\text{depth},c)$ sits at $+0.22$ to $+0.44$ on six scenes but
collapses to $+0.04$ on \texttt{tabletop\_separated} --- the same
scene where depth versus $\delta$-share inverts sign
(\S\ref{sup:structure}), and plausibly for the same reason: the
partition forces a family of long detour paths whose depth does not
coincide with the transport that drives cost elsewhere. (This is not
an instance of the depth/purity sign structure of main-paper \S6.2,
whose negative-$\rho$ scenes are \texttt{snooker} and
\texttt{ring\_caustic}, where depth--cost remains positive.) Depth is
the component whose relationship with cost is most
geometry-sensitive; any predictor using it as a cost proxy must do so
per scene.

\paragraph{Reading the table.}
A naive reading is that the descriptor vector tracks cost weakly. The
closer reading is that the continuous-cost signal itself is locally
degenerate on the hardest pixels, sign-unstable on the mixture
component and geometry-varying on the depth component. The mechanism
partition sits ahead of all three: it separates the saturating pixels
into their own buckets before averaging, and need not correlate with
cost to be informative --- the bucket identity is the signal.

% -------------------------------------------------------------------------
\section{Within-Bucket Heterogeneity of \texttt{direct}}
\label{sup:direct}

By construction the \texttt{direct} bucket admits both Lambertian
single-bounce transport and single-bounce glossy highlights, and the
two differ where they coexist. Splitting \texttt{direct} by the
$\gamma$ sidefield (threshold $0.5$; a \emph{proxy} for the
end-vertex lobe, not a definition --- fixed in the
floor-experiment registration before the analysis ran) on
\texttt{kitchen\_counter} isolates a glossy-lit sub-population of
$1{,}960$ pixels ($4.7\%$ of the bucket) with twice the median
time-to-quality ($64$ vs $32$~SPP), a $12\times$ censored fraction
($3.7\%$ vs $0.3\%$), and half the $\hat\sigma^2$ split-half
reliability ($0.29$ vs $0.60$) of its Lambertian complement ---
quantified heterogeneity inside the cheapest bucket. The other tested
scenes carry essentially no such sub-population ($13$ pixels on
\texttt{snooker}, $8$ on \texttt{veach-mis}); on \texttt{veach-mis}
the glossy-mediated energy instead lands in \texttt{diffuse-indirect}
and is carried by $\gamma$ (\S\ref{sup:sentinels-mis}). An explicit
lobe sub-split of \texttt{direct} would resolve this axis without
disturbing the main partition; we do not re-partition, since every
downstream number would move for a $\le5\%$ sub-population.

% -------------------------------------------------------------------------
\section{Per-Bucket Estimator-Perturbation Robustness}
\label{sup:robust}

\begin{figure}[!htbp]
  \centering
  \includegraphics[width=0.73\columnwidth]{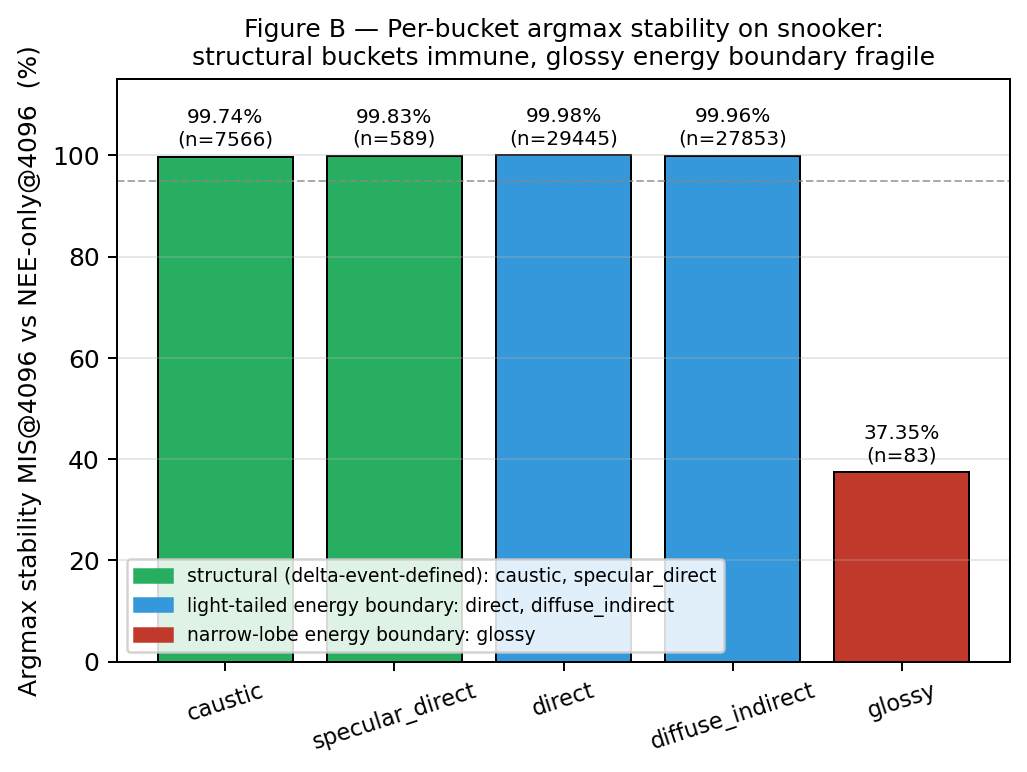}
  \caption{Per-bucket argmax stability on \texttt{snooker} under
    estimator completion (MIS@4096 vs NEE-only@4096). The predicate
    buckets (\texttt{delta-mediated}, \texttt{specular-direct}) and the
    light-tailed energy boundaries (\texttt{direct},
    \texttt{diffuse-indirect}) are all stable above $99.7\%$; only the
    narrow-lobe \texttt{glossy} energy boundary moves. Dashed line:
    $95\%$.}
  \label{sfig:per-bucket-robust}
\end{figure}

Figure~\ref{sfig:per-bucket-robust} resolves main-paper Table~4 per
bucket on \texttt{snooker}. Two implementation checks accompany the
experiment. First, on the five scenes without a glossy population ---
four of which carry substantial delta-mediated content, up to
\texttt{ring\_caustic}'s $6{,}949$ pixels --- the labelling is
unchanged \emph{to the pixel}: reweighting a bucket whose membership
is a $\delta$-event predicate cannot flip it. Second, every flip that
does occur on the remaining two scenes lies on the energy boundaries
the restricted MIS actually reweights --- overwhelmingly
\texttt{diffuse-indirect}$\leftrightarrow$\texttt{glossy} exchanges
($401$ of the $403$ \texttt{diffuse-indirect} departures on
\texttt{kitchen\_counter} go to \texttt{glossy}; the residue trades
with \texttt{delta-mediated}) --- and the single
\texttt{specular-direct}$\to$\texttt{delta-mediated} transition per
scene is a known artefact of an upper-bound test rather than a
classification ambiguity. \texttt{delta-mediated} agrees at $99.2\%$
or better on every scene.

% -------------------------------------------------------------------------
\section{Allocation Experiment: Full Control Set}
\label{sup:alloc}

The allocation experiment of main-paper \S7 was executed under a
written protocol with hypotheses, metrics and decision rules fixed
before the runs. The first execution required three disclosed
corrections, none of which altered thresholds, metrics, scenes or
treatment definitions: the allocation exponent was fixed to Neyman
$n\propto\sqrt{D}$ for every strategy alike; strict equal totals were
enforced by construction; and a shrink-placebo control absent from the
original design was added. The robust-variance incumbent and its
hypotheses were added by a protocol amendment with estimator
definition, block-count rule and decision criteria fixed before any
robust-variance run was executed; the amendment and its timestamps are
retained for inclusion in the public artifact, kept separate from the
original registration.

\paragraph{The per-pixel sample floor ($f{=}4$).}
One configuration choice is disclosed rather than pre-registered: a
per-pixel sample floor of $f{=}4$, applied to every strategy with the
floor's budget deducted from the Neyman share, first appeared as a
control in the budget sweeps, and we considered promoting it to the
shared base configuration. The selection rule was fixed before
inspecting its inputs and does not depend on the vs-uniform outcomes:
$f{=}4$ becomes primary only if it reduces bucket-floor's paired win
rate on no scene and preserves the \texttt{cubes\_dof} no-op. The
verdict: the no-op holds exactly, but the win rate drops in seven
scene--budget cells (e.g.\ \texttt{kitchen\_counter}@4: $9/10\to6/10$)
--- the two floors partially overlap in what they repair, so the
mechanism floor's marginal gain shrinks once a sample floor absorbs
the pilot's worst starvation. $f{=}1$ therefore remains primary and
$f{=}4$ is reported as the control
(Table~\ref{stab:downstream-f4} at $N_0{=}4$; the remaining budgets
in the artifact). The control table also answers the
question it invites --- is a one-line sample floor simply enough on its
own? No: on top of the shared floor, bucket-floor retains a positive
mean gain over observed-var on every one of the six scenes with
heavy-tailed buckets at $N_0{=}4$ (paired MSE wins $6$--$10/10$), and
still degenerates to the incumbent exactly where those buckets are
absent. The sample floor repairs starvation; the mechanism floor adds
information the sample floor does not have.

\begin{table}[!htbp]
\centering
\small
\setlength{\tabcolsep}{3.2pt}
\caption{The $f{=}4$ shared-sample-floor control configuration at
pilot $N_0{=}4$ (same protocol and pool as main-paper Table~11; the
floor's $4$ SPP per pixel is deducted from the Neyman share, totals
unchanged). The floor repairs much of the generic-starvation
component --- and only it: \texttt{ring\_caustic}'s deficit closes to
within seed noise ($+0.02$~dB against a $\pm2.2$ seed std),
\texttt{glass\_caustic} and \texttt{kitchen\_counter} remain below
uniform, and \texttt{cubes\_dof}'s no-op is exact. Bold: observed-var
at or below uniform.}
\label{stab:downstream-f4}
\begin{tabular}{@{}lcccccc@{}}
\hline
scene & uniform & obs-var & b-floor & oracle & MSE & p99 \\
\hline
\texttt{glass\_c.} & $\mathbf{30.79}$ & $30.29$ & $30.55$ & $36.69$ & 10/10 & 5/10 \\
\texttt{kitchen}   & $\mathbf{33.71}$ & $33.00$ & $33.18$ & $41.16$ & 6/10 & 10/10 \\
\texttt{snooker}   & $34.97$ & $37.95$ & $38.28$ & $42.46$ & 9/10 & 10/10 \\
\texttt{ring\_c.}  & $58.30$ & $58.32$ & $58.56$ & $67.09$ & 8/10 & 9/10 \\
\texttt{t.\_mixed} & $37.57$ & $37.90$ & $38.25$ & $44.39$ & 10/10 & 2/10 \\
\texttt{t.\_sep.}  & $38.04$ & $38.53$ & $38.76$ & $40.89$ & 10/10 & 5/10 \\
\texttt{cubes}     & $\mathbf{29.51}$ & $28.57$ & $28.57$ & $29.72$ & --- & --- \\
\hline
\end{tabular}
\end{table}

\paragraph{Companion budgets.}
Table~\ref{stab:downstream-1664} completes the primary
($f{=}1$) configuration at the two larger pilot budgets; gains
shrink as the pilot improves, as main-paper \S7.3 reports.

\begin{table}[!htbp]
\centering
\small
\setlength{\tabcolsep}{3.2pt}
\caption{Primary configuration at pilot $N_0{=}16$ (top) and
$N_0{=}64$ (bottom): PSNR (dB, mean over 10 disjoint draws) and
bucket-floor's per-draw win rates against observed-var on paired MSE
and p99. Bold: observed-var at or below uniform.}
\label{stab:downstream-1664}
\begin{tabular}{@{}lcccccc@{}}
\hline
scene & uniform & obs-var & b-floor & oracle & MSE & p99 \\
\hline
\multicolumn{7}{@{}l}{$N_0{=}16$}\\
\texttt{glass\_c.} & $\mathbf{30.79}$ & $30.59$ & $30.66$ & $36.64$ & 8/10 & 9/10 \\
\texttt{kitchen}   & $33.71$ & $34.14$ & $34.34$ & $41.18$ & 8/10 & 10/10 \\
\texttt{snooker}   & $34.97$ & $39.35$ & $39.57$ & $42.46$ & 8/10 & 10/10 \\
\texttt{ring\_c.}  & $58.30$ & $62.24$ & $62.87$ & $67.09$ & 10/10 & 5/10 \\
\texttt{t.\_mixed} & $37.57$ & $39.14$ & $39.17$ & $44.39$ & 6/10 & 1/10 \\
\texttt{t.\_sep.}  & $38.04$ & $39.60$ & $39.67$ & $40.85$ & 9/10 & 4/10 \\
\texttt{cubes}     & $\mathbf{29.51}$ & $29.37$ & $29.37$ & $29.71$ & --- & --- \\
\hline
\multicolumn{7}{@{}l}{$N_0{=}64$}\\
\texttt{glass\_c.} & $\mathbf{30.79}$ & $30.47$ & $30.45$ & $36.64$ & 3/10 & 3/10 \\
\texttt{kitchen}   & $33.71$ & $35.26$ & $35.24$ & $41.18$ & 4/10 & 10/10 \\
\texttt{snooker}   & $34.97$ & $39.88$ & $39.93$ & $42.46$ & 4/10 & 6/10 \\
\texttt{ring\_c.}  & $58.30$ & $62.98$ & $63.22$ & $67.09$ & 8/10 & 5/10 \\
\texttt{t.\_mixed} & $37.57$ & $39.46$ & $39.46$ & $44.39$ & 4/10 & 3/10 \\
\texttt{t.\_sep.}  & $38.04$ & $39.90$ & $39.90$ & $40.85$ & 4/10 & 1/10 \\
\texttt{cubes}     & $29.51$ & $29.66$ & $29.66$ & $29.71$ & --- & --- \\
\hline
\end{tabular}
\end{table}

\paragraph{Separating the two failure components experimentally.}
Three single-variable sweeps separate the generic-noise and
heavy-tail components of the pilot's failure. On the three diagnostic
scenes chosen to represent the pure small-sample case, the pure
heavy-tail case and the boundary case (\texttt{cubes\_dof},
\texttt{glass\_caustic}, \texttt{tabletop\_separated}), every gap is
unchanged when the test is repeated at $512^2$, ruling out a
resolution artefact. The small-pilot deficits \emph{amplify} with the
budget being allocated ($-0.97\to-4.39$~dB on
\texttt{kitchen\_counter} at $N_0{=}4$ as the extra budget grows from
$32$ to $128$~SPP, while the positive gaps compress ---
\texttt{snooker}@64 $+4.90\to+1.77$~dB; one near-zero case,
\texttt{tabletop\_mixed}@4 at $+0.10$, slips just below zero): a noisy
pilot does more damage the more samples it is trusted with, so the
deficit is a liability that scales with commitment rather than a
fixed cost of small pilots. And the per-pixel sample floor of $4$
repairs the small-sample component almost entirely
(\texttt{ring\_caustic}@4: $-1.13\to+0.02$~dB; \texttt{cubes\_dof}@4:
$-2.22\to-0.94$~dB) yet leaves \texttt{glass\_caustic} below uniform
at every pilot budget, without touching bucket-floor's gain ($+0.26$
vs $+0.28$~dB). The first component is a starvation pathology a dumb
floor can fix; the second is a mis-ranking pathology it cannot. The
\texttt{glass\_caustic} deficit persists at $N_0{=}16$ ($-0.20$~dB)
and $64$ ($-0.32$~dB); at a $4\times$ allocation budget the
$N_0{=}64$ deficit narrows to $-0.04$~dB. \texttt{kitchen\_counter}
and \texttt{ring\_caustic} flip positive by $N_0{=}16$ ($+0.43$ and
$+3.94$~dB over uniform).

\paragraph{Ablations disqualified by our own controls.}
Two further treatments were run and are reported as ablations, not
methods: cross-scene transfer of per-bucket scale factors fails the
no-harm criterion under leave-one-scene-out transfer, and
reliability-weighted within-bucket shrinkage has gains largely
reproduced by a bucket-free global-shrinkage placebo. That
global-shrinkage placebo --- a fallback-by-regularisation toward
uniform --- underperforms observed-var on all seven scenes at
$N_0{=}4$. Under the shared sample floor the shrink of bucket-floor's
gain with pilot size flattens (configuration-dependent).

\paragraph{Robust-variance incumbent.}
$\hat\sigma^2_{\mathrm{MoM}}$ is the median-of-means variance over
$k=\lceil\sqrt{N_0}\rceil$ contiguous equal blocks of the pilot frames
in render order (per-block unbiased variance, median across blocks;
trailing remainder discarded and counted). Block-count sensitivity
($k\in\{2,4,8\}$ at $N_0{=}16$, $\{4,8,16\}$ at $N_0{=}64$) changes no
conclusion. Any global multiplicative bias of the MoM variance cancels
under Neyman normalisation; only cross-pixel ordering acts. The
median-of-means estimate costs on the order of $0.1$~s per $256^2$
pilot against ${\sim}1$~ms for the naive sample variance --- a large
relative factor, negligible against rendering. Mean PSNR can favour
MoM without meeting the paired criterion (\texttt{glass\_caustic}@16
and \texttt{kitchen\_counter}@16: $7/10$ paired wins with gains below
one seed standard deviation --- throughout, the larger of the two
strategies' seed stds). Robust-floor qualifies under the strict criterion on
\texttt{tabletop\_mixed}@16 and \texttt{tabletop\_separated}@16/64;
its matched placebo deltas are ${\approx}0$ or negative in every
qualifying cell. The block-count mechanism sweep on
\texttt{ring\_caustic}@16 reads, for the \texttt{specular-direct}
bucket at the registered $k$: naive $0.87$ $\to$ MoM $2.6$ $\to$
MoM${+}$floor $0.70$ bucket MSE; the floor already improves the naive
pilot ($0.87\to0.67$), and the repaired level is $0.66$--$0.70$ across
the naive pilot and $k\le4$, with residual degradation to $0.895$
only at $k{=}8$, whose two-sample blocks carry a single degree of
freedom --- the same minimum-sample-size limit that rules out
$N_0{=}4$ entirely.

% -------------------------------------------------------------------------
\section{Third-Party Scenes: Complete Record}
\label{sup:sentinels}

Every first-party scene is procedurally generated by the authors.
The third-party scenes address the objection that the test
distribution's generating process was under our control. Rather than
adding scenes indiscriminately, we selected one third-party scene per
claim that author-built scenes cannot test. The four scenes were run
under a written registration (assets, protocol constants and every
prediction fixed before any bucket-instrumented render of a
third-party scene); one amendment was registered after a first-batch
outcome and is marked as such below. All predictions are reported
against their outcomes, pass or fail.

\subsection{Assets and protocol}
Scenes are from Bitterli's \emph{Rendering Resources} collection in
native Mitsuba~3 form (no format conversion): \texttt{veach-ajar}
(CC0; the classical ajar-door indirect-transport stress case),
\texttt{veach-mis} (CC0; four metal plates of graded roughness under
a graded light row, designed to stress NEE/BSDF balance),
\texttt{bathroom2} (CC-BY~3.0; an uncontrolled interior with mirror
and chrome fixtures beside glossy tiles), and \texttt{matpreview} (the
material test ball distributed with the collection, originally from
the Mitsuba repository). A pre-render
audit checks each against the scope of main-paper \S3.5: all emitters
are area lights except \texttt{matpreview}'s environment map, which is
replaced by an overhead area panel calibrated to the original's mean
image luminance (geometry and camera untouched). Renders use the
paper's estimator configuration (max depth 12, NEE, seed layout as in
the main-paper appendix) at $448\times256$ --- native aspect at the
paper's pixel scale --- with the identical 64-vs-4096-SPP stability
protocol. One deviation: the RGL measured-BRDF half of the taxonomy
scene could not be executed (the database's downloads are not
fetchable non-interactively); the layered-material half ran in full,
and the measured half is deferred with the pipeline in place. Total
compute for the entire suite, including the allocation experiment,
is under one GPU-hour.

\subsection{Sentinel 1: \texttt{veach-ajar}}
\label{sup:sentinels-ajar}
Six bucket-structure predictions were registered from XML inspection
alone (one area light in the adjacent room; one smooth dielectric;
four rough conductors): diffuse-indirect plurality; \texttt{direct}
confined to the door-gap wedge; a small nonzero delta-mediated
population; glossy present; emitter-direct $\approx0$; \texttt{other}
$=0$. All six hold (measured dominant populations: diffuse-indirect
$0.988$, delta-mediated $0.010$, glossy $0.002$, direct and
emitter-direct below $10^{-3}$). The registered operational
prediction was that pilot-variance allocation would sit at or below
uniform at $N_0{=}4$; the outcome far exceeds it (observed-var loses
$6.8$~dB; main-paper Table~13). The second half of the prediction ---
bucket-floor $\ge$ observed-var with the win-rate pattern of
main-paper \S7.3 --- holds only as an exact tie ($-0.004$~dB, $5/10$;
the registered win-rate pattern did not materialise), because the
scene's heavy-bucket population is ${\approx}1\%$. Under the $f{=}4$
control configuration, reported per a commitment registered before
the number was computed, the shared sample floor recovers $3.2$ of
the $6.8$~dB (observed-var $10.45\to13.61$; deficit
$-6.79\to-3.62$): about half the failure is starvation the floor
repairs, while the residual $-3.6$~dB --- still larger than any
first-party deficit --- persists in the dim-indirect Neyman-noise
regime neither floor addresses. The label's abstention diagnosis is
unchanged in both configurations. The \texttt{veach-ajar}
time-to-quality censoring measurement was registered but not executed
and is recorded as deferred.

\subsection{Sentinel 2: \texttt{veach-mis}}
\label{sup:sentinels-mis}
\begin{figure}[!htbp]
  \centering
  \includegraphics[width=\linewidth]{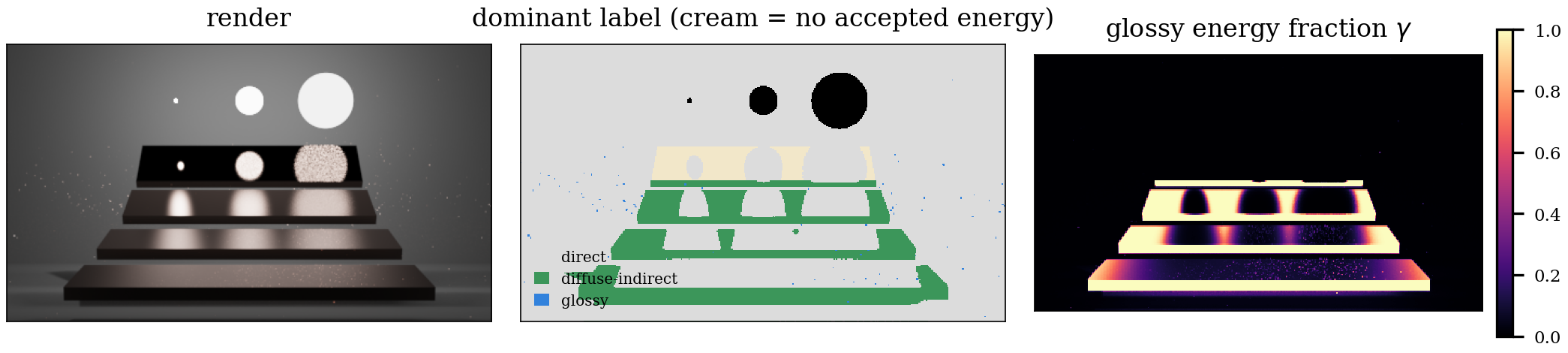}
  \caption{What the \texttt{veach-mis} plates are, per the corrected
    reading. The plate faces are \texttt{diffuse-indirect} (green)
    with glossy energy fraction $\gamma\approx1$ (right): each face
    shows the camera a glossy-mediated reflection of the diffuse
    surround, which the end-vertex rule routes to the diffuse end
    vertex. The source highlights are small \texttt{direct} ellipses;
    the near-mirror top plate has almost no accepted energy under the
    NEE-only acceptance rule (cream). Camera-side glossy mediation is
    invisible to the dominant label by construction and carried by
    $\gamma$.}
  \label{sfig:mis-mechanism}
\end{figure}

Whole-image results are unremarkable in the best sense: agreement
$0.990$, predicted $0.990$; \texttt{other} $=0\%$. The per-plate
analysis is where the failures live. The original prediction that
\texttt{glossy} would dominate the plates failed: the plate faces are
dominated by \texttt{diffuse-indirect} ($13{,}825$ of $13{,}863$
plate-face pixels; the image-wide \texttt{direct} $=0.83$ is the
diffuse surround, not the plates). What a rough plate mostly shows the
camera is a glossy-mediated reflection of the diffuse surround ---
camera $\to$ plate (G) $\to$ diffuse end-vertex --- which the
end-vertex taxonomy routes to \texttt{diffuse-indirect} by design
(Fig.~\ref{sfig:mis-mechanism}); the plate's own source highlights
(single-bounce G-end-vertex NEE events, \texttt{direct}) are small
ellipses; and the near-mirror plate's accepted energy is further
suppressed by the NEE-only acceptance rule. This reading corrects an
earlier one of our own, caught by a cross-check during the follow-up
direct-bucket analysis. The amendment registered three sharpened
predictions: (i)~plate-face glossy energy fraction $0.94$--$1.00$
against ${\approx}0.03$ elsewhere --- \textbf{pass}; (ii)~per-plate
agreement monotone in lobe width --- \textbf{fail}
($0.989/0.988/0.980/0.994$ from narrowest to widest,
$\hat\alpha\in[0.0007,0.08]$; the narrow-lobe instability of the
glossy \emph{bucket} does not transfer to these
\texttt{diffuse-indirect}-labelled pixels); (iii)~restricted-MIS
label changes concentrate on the narrowest plate --- \textbf{fail}
($4/1/8/4$ flips across plate faces; $17$ of the image's $150$ flips
lie on them at all; consistent in retrospect, since the restricted
completion touches only \texttt{glossy} and \texttt{delta-mediated}
and the plate-face pixels are \texttt{diffuse-indirect}). Both
failures are the predictor's misreading of the taxonomy's scope, not
label instability --- the labels moved nowhere unexpected --- and are
retained in the record as registered.

\subsection{Sentinel 3: \texttt{bathroom2}}
The declared separation measures were computed on the scene first
(mutual visibility $0.327$ --- higher than any tested scene --- at
$0.34$~m, between the \texttt{tabletop\_mixed} and
\texttt{kitchen\_counter} anchors), the sign prediction
$0<\rho_{C\text{--}G}<+0.441$ was registered, and only then was the
scene rendered: $\rho_{C\text{--}G}=+0.251$ --- \textbf{pass}, sign
and interval as registered (main-paper \S6.1).

\subsection{Sentinel 4: \texttt{matpreview}}
Three registered questions on the \texttt{principled} BSDF with
clearcoat swept over $\{0,0.25,0.5,1.0\}$ (a smooth coating layered
over a rough base). Does the NEE end-vertex rule remain well-defined
at every coat value? Yes: the BSDF advertises Diffuse and Glossy
components throughout, so the D-if-any-diffuse-flag rule routes NEE
events to \textbf{D} deterministically. Do clearcoat-bearing pixels
land under D-routed labels and stay there? Yes: \texttt{direct}
$0.88$, \texttt{diffuse-indirect} $0.12$; agreement $0.989$ at every
coat value; sweeping the coat from $0$ to $1$ moves $0.2\%$ of
pixels. Does \texttt{other} stay at zero? Yes, exactly. The glossy
energy fraction $\gamma$ responds monotonically to the sweep (mean
$0.040\to0.047$, $+16\%$; 90th percentile $0.166\to0.196$) while the
dominant label holds still --- the division of labour between
discrete label and continuous mixture as designed.

\subsection{Scoreboard}
Of the nineteen registered predictions scored, fifteen passed, one
held as a boundary case (the \texttt{veach-ajar} tie, consistent with
the descriptor's intended scope), and three failed --- the
original plate-dominance prediction and two of the amendment's
sharpened successors, all three the predictor's misuse of the
taxonomy's scope. Two further registered items are disclosed rather
than scored: the superseded original wording of one plate prediction
(its sharpened successor is scored in its place), and the
\texttt{veach-ajar} time-to-quality censoring measurement, registered
but not executed. The per-scene stability predictions
(main-paper Table~5) were issued from each scene's own pilot via the
frozen margin/purity curves before the reference render existed, and
land within $0.018$ on \texttt{veach-ajar} and within $0.005$ on the
other two; the curves' leave-one-scene-out error on the first-party
scenes is at most $0.061$ (purity) and $0.079$ (margin), the worst
case being \texttt{kitchen\_counter}, the most mechanism-mixed scene.
The registration document, its amendment, the frozen curves, every
per-scene prediction and outcome file, and the variant-generation
script are retained for inclusion in the public artifact
(main-paper Appendix~A).

\end{document}